\documentclass[%
reprint,
amsmath,amssymb,
aps,
prl,
floatfix,
twocolumn]{revtex4-1}

\usepackage{graphicx}
\usepackage{dcolumn}
\usepackage{bm}
\usepackage{amsmath} 
\usepackage{amssymb}
\usepackage{mathrsfs}
\usepackage{siunitx}
\usepackage{xcolor}
\usepackage{tabularx}
\usepackage{microtype}
\usepackage{hyperref}
\usepackage{dblfloatfix}
\hypersetup{
    colorlinks=true,
    linkcolor=blue,
    filecolor=magenta,      
    urlcolor=cyan,
}

\begin{document}

\onecolumngrid
\begin{center}

\textbf{\large Research Article Summary: \\ Inferring change points in the COVID-19 spreading \\ reveals the effectiveness of interventions}
\vspace{-0.15cm}

\end{center}

\begin{center}
Jonas Dehning,$^{1\ast}$ 
Johannes Zierenberg,$^{1\ast}$ 
F.~Paul Spitzner,$^{1\ast}$\\
Michael Wibral,$^{2}$
Joao Pinheiro Neto,$^{1}$
Michael Wilczek,$^{1\ast}$%
Viola Priesemann$^{1\ast}$ \\[.5em]
\normalsize{$^1$ Max Planck Institute for Dynamics and Self-Organization, G\"ottingen, Germany,}\\
\normalsize{$^2$ Campus Institute for Dynamics of Biological Networks, University G\"ottingen, Germany}\\[.5em]
\normalsize{$^\ast$ Authors contributed equally. Correspondence should be addressed to viola.priesemann@ds.mpg.de.}\\
\date{\today}
\end{center}
\vspace{-0.15cm}

\twocolumngrid
\setcounter{figure}{-1}

\paragraph{Introduction.} When faced with the outbreak of a novel epidemic like COVID-19, rapid response measures are required by individuals as well as by society as a whole to mitigate the spread of the virus. During this initial, time-critical period, neither the central epidemiological parameters, nor the effectiveness of interventions like cancellation of public events, school closings, and social distancing are known.


\paragraph{Rationale.} As one of the key epidemiological parameters, we infer the spreading rate $\lambda$ from confirmed COVID-19 case numbers at the example of Germany by combining Bayesian inference based on Markov-Chain Monte-Carlo sampling with a class of SIR (Susceptible-Infected-Recovered) compartmental models from epidemiology. Our analysis characterizes the temporal change of the spreading rate and, importantly, allows us to identify potential change points and to provide short-term forecast scenarios based on various degrees of social distancing. A detailed description is provided in the accompanying paper, and the models, inference, and predictions are available on  \href{https://github.com/Priesemann-Group/covid19_inference_forecast}{github}.
While we apply it to Germany, our approach can be readily adapted to other countries or regions.


\paragraph{Results.}

In Germany, interventions to contain the outbreak were implemented in three steps over three weeks: Around March 9, large public events like soccer matches were cancelled. On March 16, schools and childcare facilities as well as many non-essential stores were closed. One week later, on March 23, a far-reaching contact ban (``Kontaktsperre"), which included the prohibition of even small public gatherings as well as the further closing of restaurants and non-essential stores, was imposed by the government authorities.

From the observed case numbers of COVID-19, we can quantify the impact of these measures on the disease spread (Fig.~0). Based on our analysis, which includes data until April 21, we have evidence of three change points: the first changed the spreading rate from $\lambda_0=0.43$ (95~\% credible interval (CI: $[0.35,0.51]$)) to $\lambda_1=0.25$ (CI: $[0.20,0.30]$), and occurred around March 6 (CI: March 2 to March 9); the second change point resulted in $\lambda_2=0.15$ (CI: $[0.12,0.20]$), and occurred around March 15 (CI: March 13 to March 17). Both changes in $\lambda$ slowed the spread of the virus, but still implied exponential growth (Fig.~0, red and orange traces). To contain the disease spread, and turn from exponential growth to a decline of new cases, a further decrease in $\lambda$ was necessary. Our analysis shows that this transition has been reached by the third change point that resulted in $\lambda_3=0.09$ (CI: $[0.06,0.12]$) around March 23 (CI: March 20 to March 25).

With this third change point, $\lambda$ transitioned below the critical value where the spreading rate $\lambda$ balances the recovery rate $\mu$, i.e.~the effective growth rate $\lambda^* = \lambda-\mu \approx 0$ (Fig.~0, gray traces). Importantly, $\lambda^\ast=0$ presents the watershed between exponential growth or decay. Given the delay of approximately two weeks between an intervention and first inference of the induced changes in $\lambda^\ast$, future interventions such as lifting restrictions warrant careful consideration.

Our detailed analysis shows that, \textit{in the current phase}, reliable short- and long-term forecasts are very difficult as they critically hinge on how the epidemiological parameters change in response to interventions:  In Fig.~0 already the three example scenarios quickly diverge from each other, and consequently span a considerable range of future case numbers. Thus, any uncertainty on the magnitude of our social distancing in the past two weeks can have a major impact on the case numbers in the next two weeks. Beyond two weeks, the case numbers depend on our future behavior, for which we have to make explicit assumptions. In the main paper we illustrate how the precise magnitude and timing of potential change points impact the forecast of case numbers (Fig.~2). 


\paragraph{Conclusions.}

We developed a Bayesian framework to infer central epidemiological parameters and the timing and magnitude of intervention effects. Thereby, the efficiency of political and individual intervention measures for social distancing and containment can be assessed in a timely manner. We find evidence for a successive decrease of the spreading rate in Germany around March 6 and around March 15, which significantly reduced the magnitude of exponential growth, but was not sufficient to turn growth into decay. Our analysis also shows that a further decrease of the spreading rate occurred around March 23, turning exponential growth into decay. Future interventions and lifting of restrictions can be modeled as additional change points, enabling short-term forecasts for case numbers. In general, our analysis code may help to infer the efficiency of measures taken in other countries and inform policy makers about tightening, loosening and selecting appropriate rules for containment.

\begin{figure*}[t]
    \begin{center}
	\includegraphics[width=0.75\linewidth]{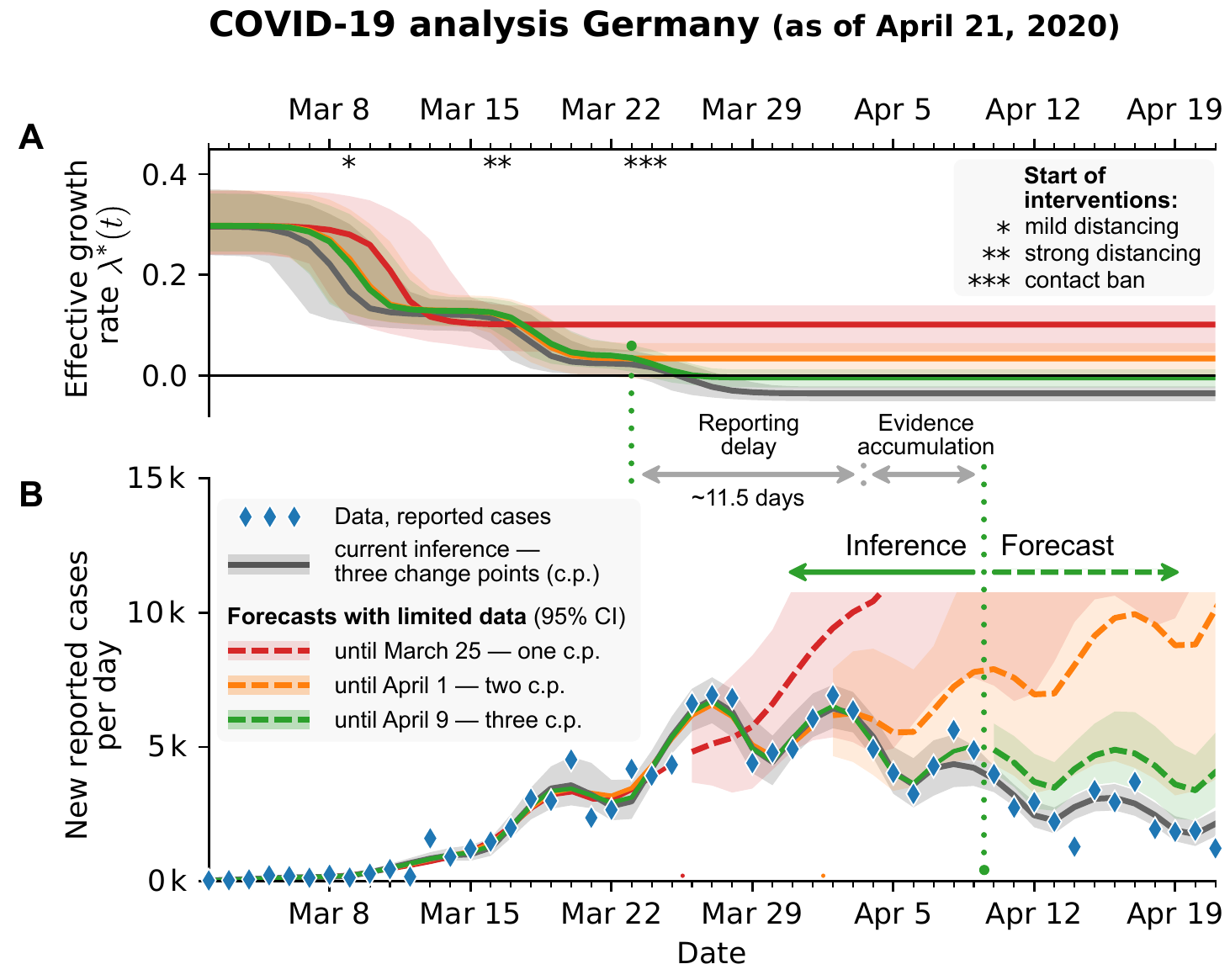}
	\end{center}
	\caption{%
	 Bayesian analysis of the German COVID-19 data (blue diamonds) reveals three change points that match the timing of publicly announced interventions.
	\textbf{A:} The inferred effective growth rate (difference between spreading and recovery rate, $\lambda^*= \lambda -\mu$) for an SIR model with weekly reporting modulation and reporting delay that includes scenarios with one, two or three change points (red, orange, green; fitted to case reports until March 25, April 1 and April 9, respectively).
    The timing of the inferred change points in growth rate is consistent with the timing of German governmental interventions (depicted as $*$, $**$, and $***$). \textbf{B:} Comparing inferred models with the actual new reported cases per day reveals the effectiveness of governmental interventions. After the first two interventions, the number of new cases still grew exponentially (red, orange); only after the third intervention did the number of new cases start to saturate (green) or even to decline (gray, data until April 21).
    This illustrates that the future development strongly depends on our distancing behavior. 
    Note the delay between a change point and the observation of changes in the number of new cases of almost two weeks --- a combination of reporting delay and a minimal period of evidence accumulation.
		\label{Fig:Fig0}
	}
\end{figure*}

\clearpage
\onecolumngrid

\title{Inferring change points in the COVID-19 spreading \\ reveals the effectiveness of interventions}

\author{Jonas Dehning$^{1*}$}%
\author{Johannes Zierenberg$^{1*}$}%
\author{F. Paul Spitzner$^{1*}$}
\author{Michael Wibral$^{2}$}
\author{Joao Pinheiro Neto$^{1}$}
\author{Michael Wilczek$^{1*}$}%
\author{Viola Priesemann$^{1}$}%
\email{viola.priesemann@ds.mpg.de}
\thanks{Authors contributed equally}

\affiliation{$^1$ Max Planck Institute for Dynamics and Self-Organization, G\"ottingen, Germany}
\affiliation{$^2$ Campus Institute for Dynamics of Biological Networks, University of G\"ottingen, Germany}

\date{\today}

\begin{abstract}

As COVID-19 is rapidly spreading across the globe, short-term modeling forecasts provide time-critical information for decisions on containment and mitigation strategies. A main challenge for short-term forecasts is the assessment of key epidemiological parameters and how they change when first interventions show an effect. By combining an established epidemiological model with Bayesian inference, we analyze the time dependence of the effective growth rate of new infections. Focusing on the COVID-19 spread in Germany, we detect change points in the effective growth rate that correlate well with the times of publicly announced interventions. Thereby, we can quantify the effect of interventions, and we can incorporate the corresponding change points into forecasts of future scenarios and case numbers. Our code is freely available and can be readily adapted to any country or region.

\end{abstract}

\maketitle

\onecolumngrid
\section{Introduction}
During the initial outbreak of an epidemic, reliable short-term forecasts are key to estimate required medical capacities, and to inform and advice the public and decision makers~\cite{enserink2020}. During this initial phase, three tasks are of particular importance to provide time-critical information for crisis mitigation: (1) establishing central epidemiological parameters, such as the basic reproduction number, that can be used for short-term forecasting; (2) simulating the effects of different possible interventions aimed at the mitigation of the outbreak; 
(3) estimating the actual effects of the measures taken --- to rapidly adjust them and to adapt short-term forecasts. Tackling these tasks is challenging due to the large statistical and systematic errors that are present during the initial stages of an epidemic with its low case numbers. This is further complicated by the fact that mitigation measures are taken rapidly, while the outbreak unfolds, but they take an effect only after an a priori unknown delay. To obtain reasonable parameter estimates for short-term forecasting and policy evaluation despite these complications, any prior knowledge available needs to be integrated into modeling efforts to reduce uncertainties. This includes knowledge about basic mechanisms of disease transmission, recovery, as well as preliminary estimates of epidemiological parameters from other countries, or from closely related pathogens. The integration of prior knowledge, the quantitative assessment of the remaining uncertainties about epidemiological parameters, and the principled propagation of these uncertainties into forecasts is the domain of Bayesian modeling and inference~\cite{jaynes2003,gelman2013}.

Here, we draw on an established class of models for epidemic outbreaks: The Susceptible-Infected-Recovered (SIR) model \cite{kermack1927,hethcote2000,anderson2000,grassly2008} specifies the rates with which population compartments change over time, i.e., with which susceptible people become infectious, or infectious people recover. This simple model can be formulated in terms of coupled ordinary differential equations (in mean field), which enable analytical treatment~\cite{parshani2010,harko2014} or fast evaluation (ideally suited for Bayesian inference). Accordingly, SIR-like models have been used to model epidemic spreads, from Bayesian Markov-Chain Monte Carlo (MCMC) parameter  estimation~\cite{britton2002,lourenco2017,faria2017} to
detailed scenario discussions~\cite{shulgin1998,bjornstad2002,hufnagel2004,pandey2014}. Recently, this family of models also played a dominant role in the analyses of the global corona virus (SARS-CoV-2) outbreak, from inference~\cite{li2020b,kucharski2020,lourenco2020} to scenario forecast~\cite{maier2020,bittihn2020,anderson2020,fauver2020,arenas2020,mitja2020,chang2020,bock2020} to control strategies~\cite{gros2020,zlatic2020}.

We combine the SIR model (and generalizations thereof) with Bayesian parameter inference and augment the model by a time-dependent spreading rate. The time dependence is implemented via potential change points that reflect changes in the spreading rate driven by governmental interventions. Based on three distinct measures taken in Germany, we detect three corresponding change points from reported COVID-19 case numbers. Already on April 1 we had reported evidence for the first two change points, and predicted the third one~\cite{dehning2020}. Now, with data until April 21, we have evidence for all three change points. First, the spreading rate decreased from 0.43 (with 95\% credible interval, CI $[0.35,0.51]$) to 0.25 (CI $[0.20,0.30]$), with this decrease initiated around March 6 (CI [2, 9]). This matches the cancellation of large public events such as trade fairs and soccer matches. Second, the spreading rate decreased further to 0.15 (CI $[0.12,0.20]$) initiated around March 15 (CI [13, 17]). This matches the closing of schools, childcare facilities, and non-essential stores. Third, the spreading rate decreased further to 0.09 (CI $[0.06,0.13]$) initiated around March 23 (CI [20, 25]). This corresponds well to the strict contact ban, which was announced on March 22. While the first two change points were not sufficient to switch from growth of novel cases to a decline, the third change point brought this crucial reversal.

Our framework is designed to infer the effectiveness of past measures and to explore potential future scenarios along with propagating the respective uncertainties. In the following, we demonstrate the potential impact of timing and magnitude of change points, and report our inference about the three past governmental interventions in Germany. Our framework can be readily adapted to any other country or region.
The code (already including data sources from many other countries), as well as the figures are all available on Github~\cite{dehning2020b}.


\section{Background: Inference of central epidemiological parameters and the effects of interventions}

In order to simulate the general effect of different possible interventions on the spread of COVID-19 in Germany, we first focus on the initial phase of the outbreak when no serious mitigation measures were implemented. In the absence of interventions, an epidemic outbreak can be described by SIR models with constant spreading rate (Methods). In Germany, first serious interventions occurred around March 9 and affected the case reports with an observation delay (a combination of incubation period with median 5--6 days~\cite{lauer2020}) and a test delay (time until doctor is visited plus test-evaluation time) that we assume to be both about 2--3 days.
Hence,  in order to infer central epidemiological parameters, we consider as the initial phase the time period from March 2 to March 15.  In order to simulate the effect of different possible interventions, we then model the effects of interventions as change points in the spreading rate (Methods).

\begin{figure}
\centering
\includegraphics[width=0.8\textwidth]{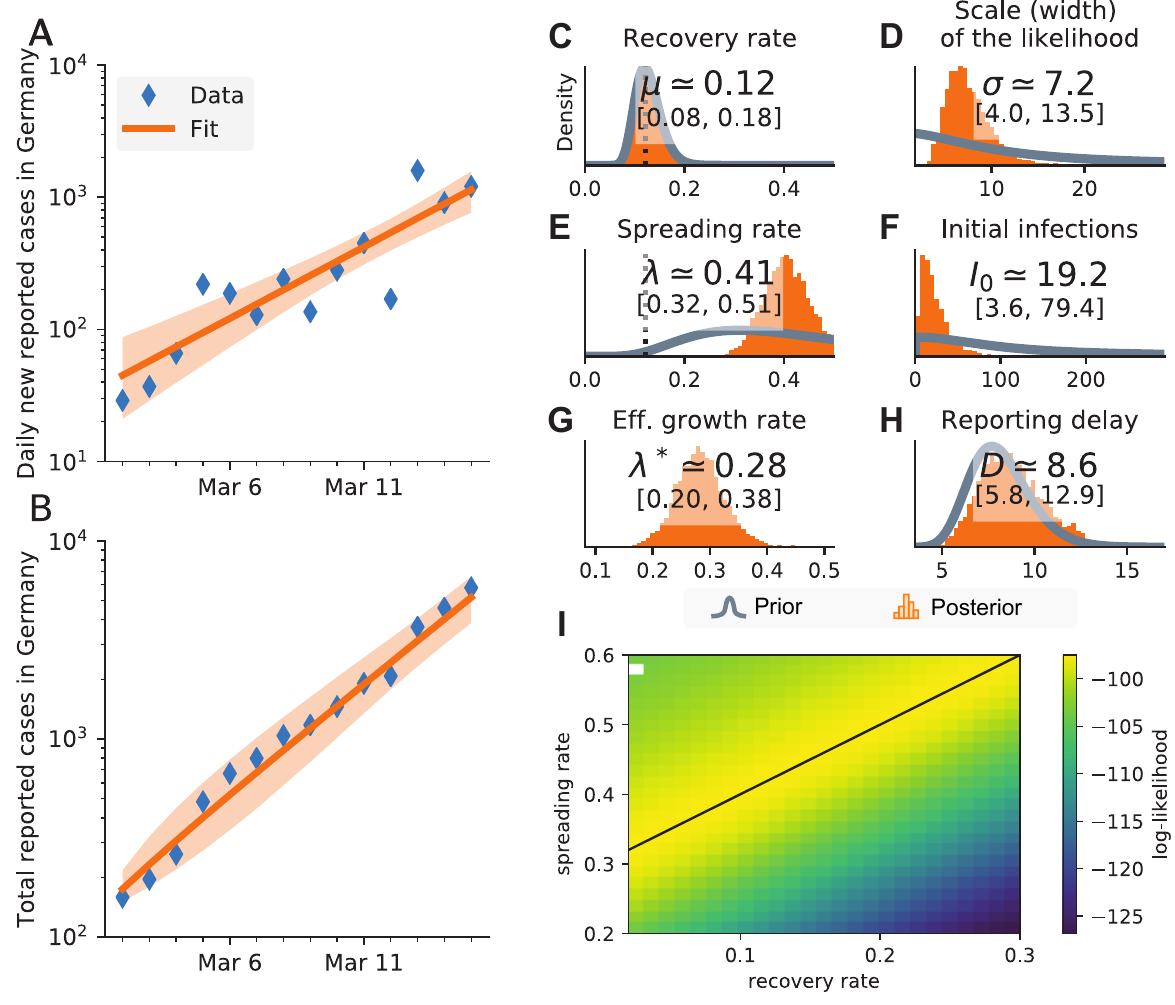}
\caption{
Inference of central epidemiological parameters of the SIR model during the initial onset period, March 2--15.
\textbf{A:}~The number of new cases and 
\textbf{B:}~the total  (cumulative) number of cases increase exponentially over time.
\textbf{C--H:} Prior (gray) and posterior (orange) distributions for all model parameters: estimated spreading rate $\lambda$, recovery rate $\mu$, reporting delay $D$ between infection date and reporting date, number of cases $I_0$ at the start of the simulation, scale-factor $\sigma$ of the width of the likelihood distribution, and the effective growth rate $\lambda^\ast=\lambda -\mu$. 
\textbf{I:} Log-likelihood distribution for different combinations of $\lambda$ and $\mu$. A linear combination of $\lambda$ and $\mu$ yields the same maximal likelihood (black line). White dot: Inference did not converge.
}
\label{fig:simple}
\end{figure}

\subsection{Bayesian inference for central epidemiological parameters during the initial phase of the outbreak} 
We perform Bayesian inference for the central epidemiological parameters of an SIR model using Markov-Chain Monte Carlo (MCMC) sampling (Fig.~\ref{fig:simple}). The central parameters are the spreading rate $\lambda$, the recovery rate $\mu$, the reporting delay $D$ and the number of initially infected people $I_0$. We chose informative priors based on available knowledge for $\lambda$, $\mu$ and $D$, and we chose uninformative priors for the remaining parameters (Methods). Also, we intentionally kept the informative priors as broad as possible such that the data would constrain the parameters (Fig.~\ref{fig:simple}).

As median estimates, we obtain for the spreading rate $\lambda=0.41$, $\mu=0.12$, $D=8.6$, and $I_0=19$ (see Fig.~\ref{fig:simple}~C--H for the posterior distributions and the 95\% credible intervals). Converted to the basic reproduction number $R_0=\lambda/\mu$, we find a median $R_0=3.4$ (CI [$2.4$, $4.7$]), which is consistent with previous reports that find median values between 2.3 and 3.3~\cite{zhang2020a,liu2020a,kucharski2020}.
Overall, the model shows good agreement with both new cases (Fig.~\ref{fig:simple}~A) and cumulative cases (Fig.~\ref{fig:simple}~B) that show the expected exponential growth (linear in log-lin plot). 
The observed data are clearly informative about $\lambda$, $I_0$ and $\sigma$ (indicated by the difference between the priors (gray line) and posteriors (histograms) in Fig.~\ref{fig:simple}~D,E,F). However, $\mu$ and $D$ are largely determined by our prior choice of parameters (histograms match gray line in Fig.~\ref{fig:simple}~C,H). This is to be expected for the initial phase of an epidemic outbreak, which is dominated by exponential growth.

In order to quantify the impact of possible interventions, we concentrate on the effective growth of active infections before and after the intervention. As long as the number of infections and recoveries are small compared to the population size, the number of active infections per day can be approximated by an exponential growth (Fig.~\ref{fig:simple}A,B) with effective growth rate $\lambda^\ast=\lambda-\mu$ (see Methods). As a consequence, $\lambda$ and $\mu$ cannot be estimated independently. This is further supported by a systematic scan of the model's log-likelihood in the $\lambda$--$\mu$ space that shows an equipotential line for the maximum likelihood (Fig.~\ref{fig:simple}~I). This strongly suggests that the growth rate $\lambda^\ast$ is the relevant free parameter with a median $\lambda^\ast = 28\%$ (Fig.~\ref{fig:simple}~G). The control parameter of the dynamics in the exponential phase is thus the (effective) growth rate: If the growth rate is larger than zero ($\lambda >\mu$), case numbers grow exponentially; if the growth rate is smaller than zero ($\lambda < \mu$), the recovery dominates and the new cases decrease. The two different dynamics (supercritical and subcritical, respectively) are separated by a critical point at $\lambda^\ast=0$ ($\lambda = \mu$)~\cite{munoz2018}. 

\subsection{Magnitude and timing of interventions matter for the mitigation of the outbreak}

\begin{figure}
    \centering
    \includegraphics[width=\textwidth]{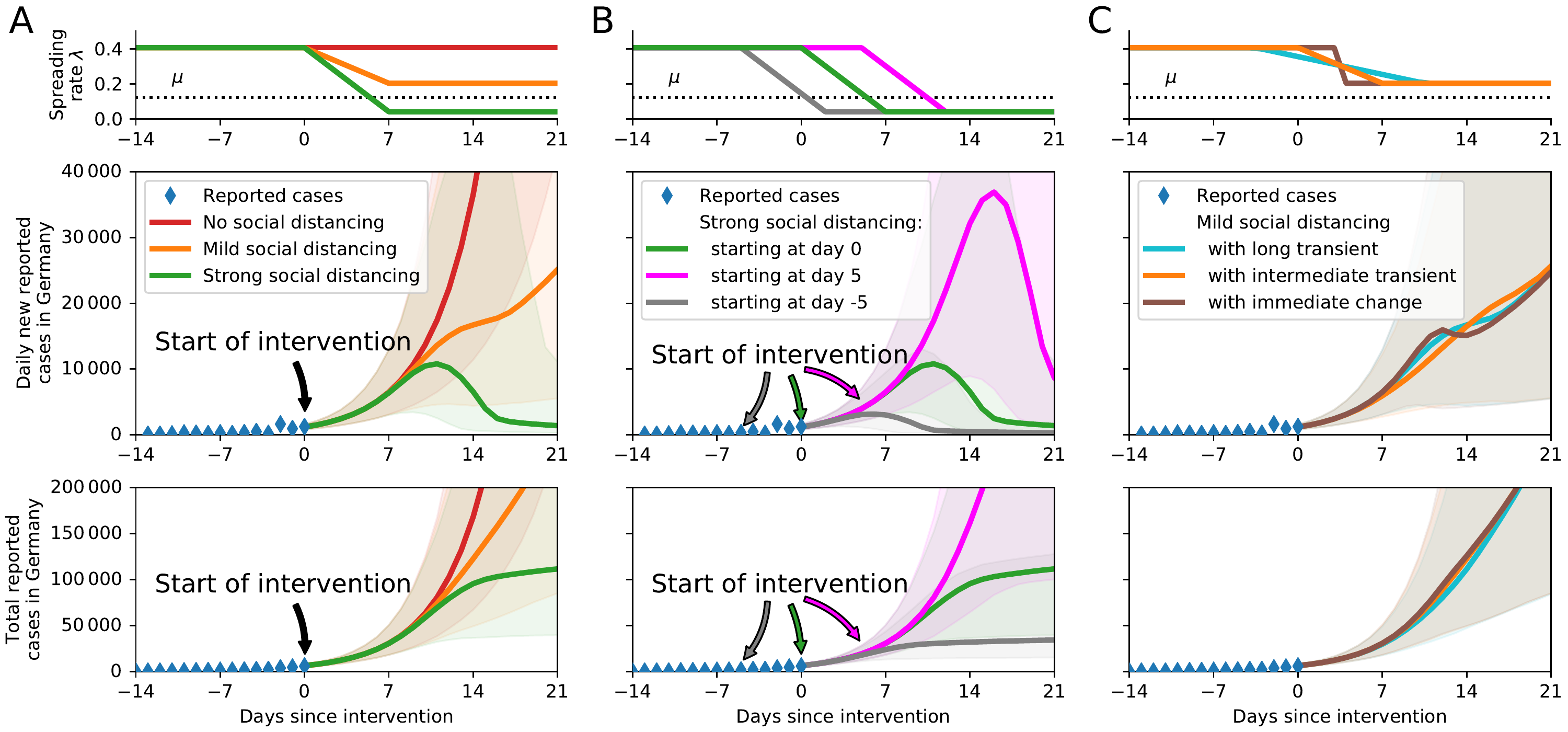}
    \caption{The timing and effectiveness of interventions strongly impact future COVID-19 cases. \textbf{A:} We assume three different scenarios for interventions starting on March 16: (I, red) no social distancing, (II, orange) mild social distancing, or (III, green) strict social distancing. \textbf{B:} Delaying the restrictions has a major impact on case numbers: strict restrictions starting on March 16 (green), five days later (magenta) or five days earlier (gray).  \textbf{C:} Comparison of the time span over which interventions ramp up to full effect. For all ramps that are \textit{centered around the same day}, the resulting case numbers are fairly similar. However, a sudden change of the spreading rate can cause a temporary decrease of daily new cases, although $\lambda > \mu$ at all times (brown).}
    \label{fig:timing}
\end{figure}

We simulate different, hypothetical interventions following the initial phase in order to show that both, the amount of change in behavior (leading to a change in spreading rate $\lambda$, Fig.~\ref{fig:timing}~A) and the exact timing of the change (Fig.~\ref{fig:timing}~B) determine the future development. 
Hypothetical interventions build on the inferred parameters from the initial phase (Fig.~\ref{fig:simple}, in particular median $\lambda_0=0.41$ and median $\mu=0.12$) and were implemented as change points in the spreading rate from the inferred $\lambda_0$ to a new value $\lambda_1$. With such a change point, we model three potential scenarios of public behavior:
\textbf{(I) No social distancing;} Public behavior is unaltered and the spread continues with the inferred rate ($\lambda_1=\lambda_0$ with median $\lambda_1=0.41>\mu$).
\textbf{(II) Mild social distancing;} The spreading rate decreases to $50\%$, ($\lambda_1 = \lambda_0\,/\,2$ with median $\lambda_1=0.21>\mu$). Although people effectively reduce the number of contacts by a factor of two in this second scenario, the total number of reported cases continues to grow alongside scenario (I) for the time period of the reporting delay $D$ (median $D=8.6$ from initial phase, see below for a more constrained estimation). Also, we still observe an exponential increase of new infections after the intervention becomes effective, because the growth rate remains positive, $\lambda^\ast_1=\lambda_1-\mu>0$. 
\textbf{ (III) Strong social distancing;} Here, the spreading rate decreases to $10\%$, ($\lambda_1 = \lambda_0\,/\,10$ with median $\lambda_1=0.04<\mu$). The assumptions here are that contacts are severely limited, but even when people stay at home as much as possible, \textit{some} contacts are still unavoidable. Even under such drastic policy changes, no effect is visible until the reporting delay $D$ is over. Thereafter, a quick decrease in daily new infections manifests within two weeks (delay plus change point duration), and the total number of cases reaches a stable plateau. Only in this last scenario a plateau is reached, because here the growth rate becomes negative, $\lambda^\ast<0$, which leads to decreasing numbers of new infections.

Furthermore, the timing of an intervention matters: Apart from the strength of an intervention, its \textit{onset} time has great impact on the total case number (Fig.~\ref{fig:timing}~B,C). For example, focusing on the strong intervention (III)
--- where a stable plateau is reached --- the effect of advancing or delaying the change point by just five days leads to more than a three-fold difference in cumulative cases.

While we find that the timing of an intervention has great effect on case numbers, the duration over which the change takes place has only minor effect --- if the intervals of change are centered around the same date.
In Fig.~\ref{fig:timing}~C we illustrate the adjustment of $\lambda_0 \to \lambda_1$ for mild social distancing with durations of 14, 7 and 1 day(s). The change point duration is a simple way to incorporate variability in individual behavior, and is not linked to the reporting delay $D$. As an interesting effect, a sudden change in the spreading rate can lead to a temporary decrease of new case numbers, despite the fact that the effective growth rate remains positive at all times.


\section{Results}

\begin{figure}
    \centering
    \centerline{
    \includegraphics[scale=.8]{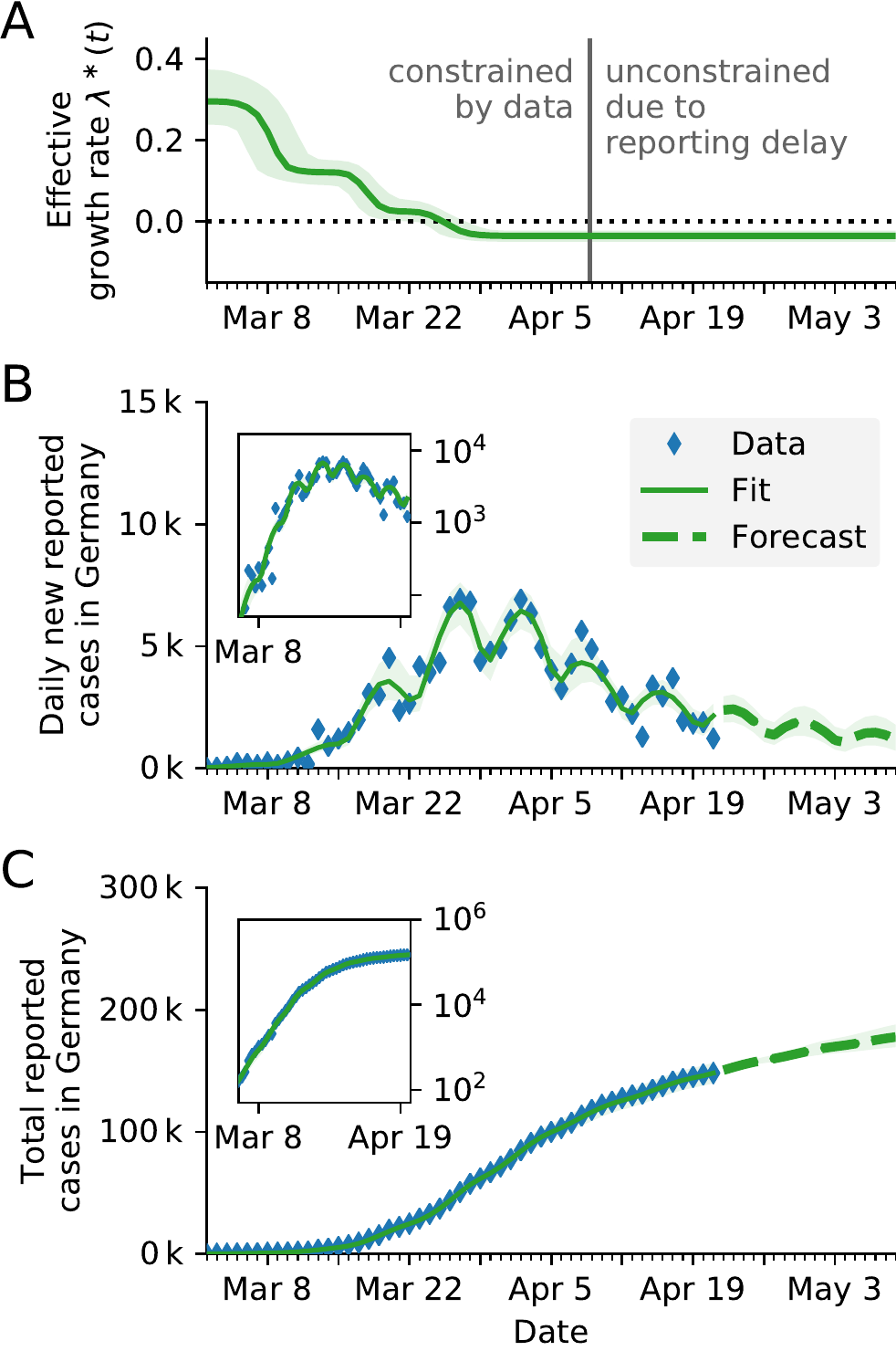}
    \hspace{2mm}
    \includegraphics[scale=.8]{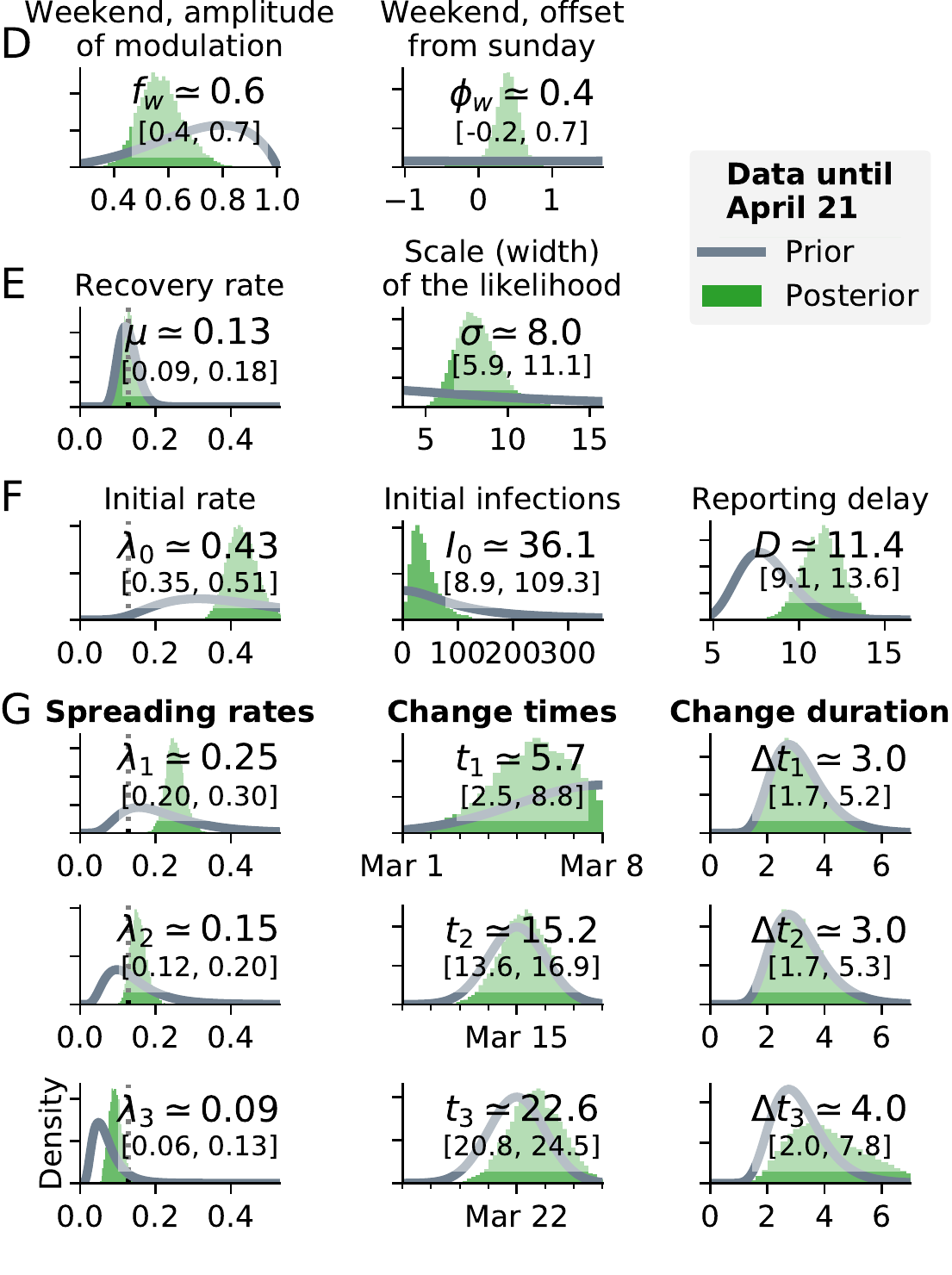}
    }
    \caption{Bayesian analysis of the German COVID-19 data (blue diamonds) until April 21 reveals three change points that are consistent with three major governmental interventions. 
    \textbf{A:}~Time-dependent model estimate of the effective spreading rate $\lambda^*(t)$.
    \textbf{B:}~Comparison of daily new reported cases and the model (green solid line for median fit with 95\% credible intervals, dashed line for median forecast with 95\% CI); \textbf{inset:} same data in log-lin scale. \textbf{C:} Comparison of total reported cases and the model (same representation as in B).
    \textbf{D--F:}~Priors (gray lines) and posteriors (green histograms) of all model parameters; inset values indicate the median and 95\% credible intervals of the posteriors. 
    For the same model with one or two change points, please see the corresponding figures in the SI (Fig.~\ref{fig:inference_1_change_point}, \ref{fig:inference_2_change_points}, Table~\ref{tab:loo_suppls}).
    %
    }
    \label{fig:inference_3_change_points}
\end{figure}

In order to model real-world data, we further refined the SIR model. Most importantly, we account for systematic variations of case reports throughout the week, in particular lower case numbers on the weekend, by explicitly modeling a weekly reporting modulation (see Methods). Indeed, model comparisons confirm that models with this correction outperform those without (see Table~\ref{tab:loo_suppls}). In the supplemental material, we further generalize our model to include an explicit incubation period (SEIR-like, Fig.~\ref{fig:inference_SEIR_not_rw}) that yields results consistent with our main model.

We incorporate the effect of governmental interventions into our models by introducing flexible change points in the spreading rate (see Methods). During the COVID-19 outbreak in Germany, governmental interventions occurred in three stages from (i) the cancellation of large events with more than 1000 participants (around March 9), through (ii) closing of schools, childcare facilities and the majority of stores (in effect March 16), to (iii) the contact ban and closing of all non-essential stores (in effect March 23). The aim of all these interventions was to reduce the (effective) growth rate $\lambda^\ast = \lambda - \mu$. As soon as the growth rate becomes negative ($\lambda^\ast<0$), the number of new confirmed infections decreases after the respective reporting delay.

Detecting change points in the spreading rate --- and quantifying the amount of change as quickly as possible --- becomes a central modeling challenge when short-term forecasts are required. To address this challenge, we assume an initial spreading rate $\lambda_0$ (the exponential growth phase, cf.~Fig.~\ref{fig:simple}) and up to three potential change points motivated by the German governmental interventions: The first change point ($\lambda_0 \to \lambda_1$) is expected around March 9 ($t_1$) as a result of the official recommendations to cancel large events. A second change point ($\lambda_1 \to \lambda_2$) is expected around March 16 ($t_2$), when schools and many stores were closed. A third change point ($\lambda_2 \to \lambda_3$) is expected around March 23 ($t_3$), when all non-essential stores were closed, and a contact ban was enacted. We model the behavioral changes that are introduced at these change points to unfold over a few days $\Delta t_i$, but the changes in duration can be partly compensated by changes in the onset time $t_i$ (see Fig.~\ref{fig:timing}~C, scenarios). 
We chose priors for all parameters based on the information available to us up to March 28 (see Methods).  In addition, we performed a sensitivity analysis by employing wider priors in the supplemental material (Figs.~\ref{fig:SIR_wider_delay},~\ref{fig:SIR_wider_days},~\ref{fig:SIR_wider_transient}, Table~\ref{tab:loo_suppls}), which yielded consistent results. On March 28, the data were already informative about the first change point, and thereby helped to inform our forecast scenarios.

The inferred parameters for the models with change points are consistent with the inferred parameters from the exponential onset phase (Figs.~\ref{fig:simple},~\ref{fig:inference_3_change_points} \& Figs.~\ref{fig:inference_1_change_point},~\ref{fig:inference_2_change_points}). In particular, all estimated $\lambda_0$-values from models with multiple change points are compatible with the value of the model without change points (during the exponential onset phase, $\lambda_0=0.41$, CI $[0.32,0.51]$, assuming a stationary $\lambda$ until March 15, Fig.~\ref{fig:simple}~E).
Also the scale factor $\sigma$ and the number of initial infections $I_0$ for the models with change points are consistent with the initial model inference during the exponential onset phase (Fig.~\ref{fig:simple}~D,F).

\subsection{The models with two or three change points fit the observed data better than those with fewer change points}
The models with three change points describe the data better than models with fewer change points, as indicated by the leave-one-out (LOO) cross-validation-based Bayesian model comparison~\cite{vehtari2017} (lowest LOO-score in Table~\ref{tab:loo}). However, the LOO-scores of the model with two and three change points differ by less than one standard error. This originates from an extended duration of the second change point in the two-change-point model, which partially captures the effect of the third intervention. As expected, the models with none or a single change point have LOO-scores that are at least one standard deviation higher (worse) than those of the best models, and we will not consider them further.

When comparing our inference based on three change points to the number of confirmed cases, we find them to largely match (Fig.~\ref{fig:inference_3_change_points}~B,C).
The dominant periodic change in the daily new reported cases (Fig.~\ref{fig:inference_3_change_points}~B) is well described by the weekday modulation. In addition to the periodic change, the daily new case numbers also reflect the fairly sudden change of the spreading rate at the change points (cf.~Figs.~\ref{fig:timing} and~\ref{fig:SIR_without_modulation} for the effect of change points without the modulation).
Most importantly, the cumulative effect of change points manifests in an overarching decay in new case numbers that is visible after April 5 and follows the third change point (with reporting delay).

\subsection{Change point detection quantifies the effect of governmental interventions on the outbreak of COVID-19 in Germany}
Ideally, detected changes can be related to specific mitigation measures, so that one gains insights into the effectiveness of different measures (Fig.~\ref{fig:inference_3_change_points}). Indeed, we found clear evidence for three change points in the posterior distributions of the model parameters: 
First, $\lambda(t)$ decreased from $\lambda_0=0.43$ (with 95\% credible interval, CI $[0.35, 0.51]$) to $\lambda_1=0.25$ (CI [0.20, 0.30]).  The date of the change point was inferred to be March 6 (CI [2, 9])]; this inferred date matches the timing of the first governmental intervention including cancellations of large events, as well as increased awareness. After this first intervention, the (effective) growth rate $\lambda^\ast(t) = \lambda(t)-\mu$ decreased by more than a factor 2, from median $\lambda_0-\mu=0.3$ to median $\lambda_1-\mu = 0.12$, given that the recovery rate was inferred as $\mu=0.13$ (CI [0.09, 0.18]). 
Second, $\lambda(t)$ decreased from $\lambda_1=0.25$ to $\lambda_2=0.15$ (CI [0.12, 0.20]), which is larger than our prior assumption. The date of the change point was inferred to be March 15 (CI [13, 17])]; this inferred date matches the timing of the second governmental intervention including closing schools and some stores. After this second intervention, the median growth rate became $\lambda^\ast(t) = \lambda_2-\mu = 0.02 \approx 0$ and is thus in the vicinity of the critical point, yet still positive. The first two interventions in Germany thereby mitigated the spread by drastically reducing the growth rate, but the spread of COVID-19 remained exponential.
Third, $\lambda(t)$ decreased from $\lambda_2=0.15$ to $\lambda_3=0.09$ (CI [0.06, 0.13]). The date of the change point was inferred to be March 23 (CI [20, 25])]; this inferred date matches the timing of the third governmental intervention including contact ban and closing of all non-essential stores. Only after this third intervention, the median (effective) growth rate, $\lambda^\ast(t)=\lambda_3-\mu = -0.03 < 0$ (CI $[-0.05, -0.02]$)], finally became negative, indicating a decrease in the number of new infections. We can thus clearly relate the change points to the governmental interventions and quantify their mitigation effect.

\begin{table}[]
    \centering
    \caption{Model comparison shows that the three-change-point model describes the data best: leave-one-out (LOO) cross-validation for main models (SIR with weekend correction) and a different number of change points. Lower LOO-scores represent a better match between model and data.}
    \begin{tabular}{l|l|l}
        Model & LOO-score & Effective number of parameters (pLOO) \\
        \hline
        zero change points & $927 \pm 9$  & $8.31$\\
        one change points & $819 \pm 16$  & $13.46$\\
        two change points & $796 \pm 17$  & $12.53$\\
        three change points & $787 \pm 17$  & $13.42$\\
    \end{tabular}
    \label{tab:loo}
\end{table}

\section{Discussion}

We presented a Bayesian approach for a timely monitoring of the effect of governmental interventions on the spread of an epidemic outbreak.
At the example of the COVID-19 outbreak in Germany, we applied this approach to infer the central epidemiological parameters and three change points in the spreading rate from the number of reported cases.
We showed that change points in the spreading rate affect the confirmed case numbers with a delay of about two weeks (median reporting delay of $D=11.4$ days plus a median change-point duration of $3$ days). Thereby, we were able to relate the inferred change points to the three major governmental interventions in Germany: We found a clear reduction of the spreading rate related to each governmental intervention (Fig.~\ref{fig:inference_3_change_points}), (i) the cancellation of large events with more than 1000 participants (around March 9), (ii) the closing of schools, childcare centers and the majority of stores (in effect March 16), and (iii) the contact ban and closing of all non-essential stores (in effect March 23).

Our results suggest that the full extent of governmental interventions was necessary to stop exponential growth. The first two governmental interventions brought a reduction of the growth rate $\lambda^\ast$ from $30\%$ to $12\%$ and down to $2\%$, respectively. However, these numbers still implied exponential growth. Only with the third intervention --- the contact ban --- we found that we have crossed the transition in new case numbers from growth to decay. However, the decay rate of about $-3\%$ (CI $[-5\%, -2\%]$) remains close to zero. Hence, even a minor increase in the spreading rate may again switch the dynamics to the unstable regime with exponential growth.  

We used a formal Bayesian model comparison in order to validate the presence of change points. Our model comparison ruled out models with fewer than two change points (Table~\ref{tab:loo},~\ref{tab:loo_suppls}). While this may seem trivial, it has important consequences for making short-term forecasts that decision makers rely on: Demonstrating and quantifying the effect of past change points can be used to formulate priors for the effects of future, similar change points. These priors help to project the effects of more recent change points into future forecasts, even when these change points are not apparent in the reported case numbers yet. Consequently, it is important to look out for and identify potential change points as early as possible to incorporate them into forecasts. 

The detection of change points and their interpretation depend crucially on an accurate estimate of the reporting delay $D$. Therefore, the validity of its estimate should be evaluated. In our model, $D$ contains at least three distinct factors: the biological incubation period (median 5--6 days)~\cite{lauer2020}, an additional delay from first symptoms to symptoms motivating a test (1--3 days) and a possible delay before a testing results come in (1--4 days). The sum of these delays seems compatible with our inferred median delay of $D=11.4$ days, especially given the wide range of reported incubation periods. 

{We chose to keep our main model comparatively simple, because of the small number of data points initially available during an epidemic outbreak.} With such a low number of data points, only a limited number of parameters can be effectively constrained. Hence, we chose to approximate a time-dependent spreading rate $\lambda(t)$ by episodes of constant spreading rates $\lambda_i$ that are separated by three change points where a transition occurs. Our results show that this main model is currently sufficient for Germany: While we introduced fairly broad priors on the spreading rates, we obtained comparably narrow posterior distributions for each spreading rate $\lambda_i$ (Fig.~\ref{fig:inference_3_change_points}). 
We additionally evaluated extensions of our main model with three change points, e.g., by explicitly taking into account the incubation period (Fig.~\ref{fig:inference_SEIR_not_rw}). 
These models yield consistent results for the three change points, and all have LOO scores within one standard error of each other.
Thus, we consider our main model to be sufficient for case numbers in Germany at present. 

{Our framework can be easily adapted to other countries and enables one to incorporate future developments.}
For other countries, or for forecasts within smaller communities (e.g. federal states or cities), additional details may become  important, such as explicit modeling of incubation time distributions~\cite{li2020b,peng2020} (i.e.~as done in Fig. \ref{fig:inference_SEIR_not_rw}), spatial heterogeneity~\cite{li2020b,bittihn2020}, isolation effects~\cite{maier2020,peng2020}, subsampling effects hiding undetected cases even beyond the reporting delay~\cite{wilting2018,chen2020}, or the age and contact structure of the population \cite{chang2020}. In countries where drastic changes in test coverage are expected, this will have to be included as well. The methodology presented here is capable in principle of incorporating such details. It also lends itself to modeling of continuous drifts in the spreading rate, e.g. reflecting reactions of the public to news coverage of a catastrophic situation, or people growing tired of mitigation measures. Such further adaptations, however, can only be performed on a per-country basis by experts with an intimate knowledge of the local situation. Our code provides a solid and extensible base for this. For Germany, several developments in the near future may have to be included in the model. First, people may have transiently changed their behavior over the Easter holidays; second, we expect a series of change points, as well as continuous drifts, with governments trying to ease and calibrate mitigation measures. Third, extensions to hierarchical models will enable regional assessments, e.g. on the level of federal states.

Even after the three major governmental interventions in Germany, effective growth rates remain close to zero and warrant careful consideration of future measures. At present, estimates of effective growth rates dropped to $-3\%$ and thereby remain close to zero -- the watershed between exponential growth or decay. Together with the delay of approximately two weeks between infection and case report this warrants caution in lifting restrictions for two reasons: First, lifting restrictions too much will quickly lead to renewed exponential growth; second, we would be effectively blind to this worsened situation for nearly two weeks in which it will develop uninhibited. This may result in unwanted growth in case numbers beyond the level that the health system can cope with -- especially when the active cases have not gone down close to zero before lifting restrictions, thus re-initiating growth from a high base level. Therefore, it is important to consider lifting restriction only when the number of active cases are so low that a two-week increase will not pose a serious threat.

\textbf{In conclusion}, the presented Bayesian approach allows to detect and quantify the effect of recent governmental interventions and -- combined with potential subsequent interventions -- to forecast future case number scenarios. Our analysis highlights the importance of precise timing and magnitude of interventions for future case numbers. It also stresses the importance of including the reporting delay $D$ between the date of infection and the date of the confirmed case  in the model. The delay $D$, together with the time required to implement interventions  causes a total delay between an intervention and its visibility in the case numbers of about two weeks for COVID-19 in Germany. This means that changes in our behavior today can only be detected in confirmed cases in two weeks. Combined with the current spreading rate that is still around zero, the inferred spreading and observation dynamics warrant an extremely careful planning of future measures.


\section{Materials and methods}

As a basis for our Bayesian inference and the forecast scenarios, we use the differential equations of the well-established SIR (Susceptible-Infected-Recovered) model. We also test the robustness of our results by means of more sophisticated models, in particular an SEIR-like model that explicitly incorporates an incubation period (Fig.~\ref{fig:inference_SEIR_not_rw}). While the SIR model-dynamics is well understood in general, here our main challenge is to estimate model parameters specifically for the COVID-19 outbreak, and to use them for forecasting. To that end, we combined a Bayesian approach --- to incorporate prior knowledge --- with Markov Chain Monte Carlo (MCMC) sampling --- to compute the posterior distribution of the parameters and to sample from it for forecasting.
Put simply, we first estimate the parameter distribution that best describes the observed situation, and then we use many samples from this parameter distribution to evolve the model equations and thus forecast future developments. 

The data used comes from the Johns Hopkins University Center for Systems Science and Engineering (JHU CSSE) dashboard~\cite{dong2020}. The JHU CSSE provides up-to-date data on COVID-19 infections, usually a few days ahead of official German sources. The exact version of the data and code is available at~\cite{dehning2020b}. Data were incorporated until April 21. Note that after this cutoff date, additional modeling of the effects of behavioral changes over the Easter holidays becomes necessary.

\subsection{Simple model: SIR model with stationary spreading rate}

We consider a time-discrete version of the standard SIR model. In short, we assume that the disease spreads at rate $\lambda$ from the infected population compartment ($I$) to the susceptible compartment ($S$), and that the infected population compartment recovers ($R$) at rate $\mu$.
This well-established model for disease spreading can be described by the following set of (deterministic) ordinary differential equations (see, e.g., Refs~\cite{hethcote2000,anderson2000,maier2020}). Within a population of size $N$,
\begin{equation}
  \begin{array}{r@{}l@{}}
    \frac{\mathrm{d}S}{\mathrm{d}t} & {}= -\lambda\frac{S I}{N} \\[\jot]
    \frac{\mathrm{d}I}{\mathrm{d}t} & {}=  \lambda\frac{S I}{N} - \mu I\\[\jot]
    \frac{\mathrm{d}R}{\mathrm{d}t} & {}=  \mu I\,.
  \end{array}
\end{equation}

As a remark, during the onset phase of an epidemic only a very small fraction of the population is infected ($I$) or recovered  ($R$), and thus $S\approx N \gg I$ such that $S/N\approx 1$. Therefore, the differential equation for the infected reduces to a simple linear equation, exhibiting an exponential growth
\begin{equation}
\frac{dI}{dt} = (\lambda - \mu) I \quad{\rm solved\,by}\quad I(t) = I(0)~e^{(\lambda-\mu)t}\,.
\end{equation}

Because our data set is discrete in time ($\Delta t=$1 day), we solve the above differential equations with a discrete time step (${\mathrm{d}I}/{\mathrm{d}t} \approx {\Delta I}/{\Delta t}$), such that

\begin{equation}
  \begin{array}{r@{}l@{}l}
        S_t - S_{t-1} &{}= -\lambda \Delta t\frac{S_{t-1}}{N} I_{t-1} &{} =: -I_t^\mathrm{new}\\[\jot]
        R_t - R_{t-1} &{}= \phantom{-\frac{S_{t-1}}{N}}\mu \Delta t I_{t-1} &{} =: \phantom{-}R_t^\mathrm{new}\\[\jot]
        I_{t} - I_{t-1} &{}= \left(\lambda\frac{S_{t-1}}{N}-\mu\right) \Delta t I_{t-1} &{} = I_{t}^\mathrm{new} - R_{t}^\mathrm{new} \, .
  \end{array}
\end{equation}

Importantly, $I_{t}$ models the number of all (currently) active infected people, while $I_t^\mathrm{new}$ is the number of new infections that will eventually be reported according to standard WHO convention. Importantly, we explicitly include a reporting delay $D$ between new infections $I_t^\mathrm{new}$ and newly reported cases ($C_t$) as 
\begin{equation}
C_t = I_{t-D}^\mathrm{new}.
\end{equation}
We begin our simulations at time $t=0$ with $I_0$ infected cases and start including real-word data of reported cases $\hat{C}_t$ from day $t>D$ (see below for a parameterization). 

In our model we do not explicitly incorporate the inflow of additional infected people by travel for two reasons. First, we implicitly model the initial surge of infections with $I_0$. Second, previous work showed that travel during the outbreak has only modest effects on the dynamics, e.g., travel restrictions in China merely delayed the exponential spread if not combined with reductions of spreading~\cite{chinazzi2020}.

\subsection*{Full model: SIR model with weekly reporting modulation and change points in spreading rate}

Our change point detection builds on a generalization of the simple SIR model with stationary spreading rate. We now assume that the spreading rate $\lambda_i$, $i=1,...,n$, may change at certain time points $t_i$ from $\lambda_{i-1}$ to $\lambda_i$, linearly over a time window of $\Delta t_i$ days. Thereby, we account for policy changes to reduce $\lambda$, which were implemented in Germany step by step. Thus, the parameters $t_i$, $\Delta t_i$, and $\lambda_i$ are added to the parameter set of the simple model above, and the differential equations are augmented by the time-varying $\lambda_i$. 

In addition, we include a weekly modulation to account for lower case reports around the weekend which subsequently accumulate during the week. To model the systematic variation of case reports during the week, we adapted the newly reported cases by a reporting fraction
\begin{equation}
  \begin{array}{r@{}l@{}l}
  C_t &{}= I_{t-D}^\mathrm{new}~(1-f(t))\,, \qquad\mathrm{with}\\[\jot]
    f(t) &{}= (1-f_w) \cdot \left(1 - \left|\sin\left(\frac{\pi}{7} t- \frac{1}{2}\Phi_w\right)\right| \right) \,,
  \end{array}
\end{equation}
where $f_w$ and $\Phi_w$ will also be constrained by the data.

\subsection*{Estimating model parameters with Bayesian MCMC}

We estimate the set of model parameters $\theta=\{\lambda_i, t_i, \mu, D, \sigma, I_0, f_w, \Phi_w\}$ using Bayesian inference with Markov-chain Monte-Carlo (MCMC). The parameter $\sigma$ is the scale factor for the width of the likelihood $P\big(\hat{C}_t \big |\theta\big)$ between observed data and model (see below). Our implementation relies on the python package PyMC3~\cite{salvatier2016} with NUTS (No-U-Turn Sampling)~\cite{hoffman2014} using multiple, independent Markov chains.
The structure of our approach  is the following:

\paragraph{Initialization of the Markov chains via  variational inference.} The posterior is approximated by Gaussian distributions ignoring correlations between parameters through automatic differentiation variational inference (ADVI)~\cite{kucukelbir2017}, which is implemented in PyMC3. From this distribution, four starting points for four chains are sampled.

\paragraph{Burn-in phase:} Each chain performs 1000 burn-in (tuning) steps using NUTS, which are not recorded. This serves as equilibration in order to sample from an equilibrium distribution in the next step.

\paragraph{Sampling phase:} Each chain performs 4000 steps, which are used to approximate the posterior distribution. To ensure that the Chains are equilibrated and sampled from the whole posterior distribution (ergodicity), we verified that the R-hat statistic is below 1.05, which is implemented in PyMC3. The rank normalized R-hat diagnostic tests for lack of convergence by comparing the variances within chains and between chains: For identical within-chain and between-chain variances R-hat becomes 1, indicating convergence. For well-converged chains the resulting samples will describe the real-world data well, so that consistent forecasts are possible in the forecast phase.

\paragraph{Forecast using Monte Carlo samples.} For the forecast, we  take all samples from the MCMC step and continue time integration according to different forecast scenarios explained below. Note that the overall procedure yields an ensemble of forecasts --- as opposed to a single forecast that would be solely based on one set of (previously optimized) parameters.

\paragraph{MCMC sampling details} Each MCMC step requires to propose a new set of parameters $\theta$, to generate a (fully deterministic) time series of new infected cases $C(\theta)=\left\{C_t(\theta)\right\}$ of the same length as the observed real-world data $\hat{C}=\left\{\hat{C}_t\right\}$, and to accept or reject $\theta$. In our case, the NUTS implementation (in PyMC3) first proposes a new set of parameters $\theta$ based on an advanced gradient-based algorithm and subsequently accepts or rejects it such that the resulting samples reflect the posterior distribution 
$$
    p(\theta | \hat{C})\propto p(\hat{C}|\theta)p(\theta),
$$
where $p(\hat{C}|\theta)$ is the likelihood for the data given the parameters and $p(\theta)$ is the prior distribution of the parameters (see below). The likelihood quantifies the similarity between model outcome and the available real-world time series. Here, the likelihood is the product over local likelihoods

    $$ p\big(\hat{C}_t \big |\theta\big) 	\sim \rm{StudentT}_{\nu = 4} \left( 
        \rm{mean} = C_t(\theta), \,
        \rm{width} = \sigma \sqrt{C_t(\theta)}
        \right) \,. $$
quantifying the similarity between the model outcome for one time point $t$, $C_t(\theta)$, and the corresponding real-world data point $\hat{C}_t$. We chose the Student's t-distribution because it resembles a Gaussian distribution around the mean but features heavy tails, which make the MCMC more robust with respect to outliers~\cite{lange1989}, and thus reporting noise. The case-number-dependent width is motivated by observation noise through random subsampling~\cite{wilting2018}, resulting in a variance proportional to the mean. Our likelihood neglects any noise in the dynamic process, as the SIR model is deterministic, but could be in principle extended to incorporate typical demographic noise from stochastic spreading dynamics~\cite{disanto2017,munoz2018}.

\subsection*{Priors that constrain model parameters}

As short-term forecasts are time-critical at the onset of an epidemic, the available real-world data is typically not informative enough to identify all free parameters, or to empirically find their underlying distributions. We therefore chose informative priors on initial model parameters where possible and complemented them with uninformative priors otherwise. Our choices are summarized in Tab.~\ref{tab:prior_simple} for the simple model, i.e.\,the SIR model with stationary spreading rate for the exponential onset phase, and in Tab.~\ref{tab:prior_full} for the full model with change points, and are discussed in the following.

\paragraph{Priors for the simple model (Table~\ref{tab:prior_simple}):}

\begin{table}[h]
    \centering
    \caption{Priors on the model parameters for the SIR model with stationary spreading rate.}
    \begin{tabular}{r|c|l}
        Parameter & Variable & Prior distribution \\
        \hline
        Spreading rate & $\lambda$ & $\rm{LogNormal}(\log(0.4), 0.5)$\\  
        Recovery rate &$\mu$ & $\rm{LogNormal}(\log(1/8), 0.2)$\\
        Reporting delay &$D$ & $\rm{LogNormal}(\log(8), 0.2)$\\
        \hline
        Initially infected& $I_0$ & $\rm{HalfCauchy}(100)$\\
        Scale factor & $\sigma$ & $\rm{HalfCauchy}(10)$\\
    \end{tabular}
    \label{tab:prior_simple}
\end{table}

In order to constrain our simple model, an SIR model with stationary spreading rate for the exponential onset phase, we chose the following informative priors. Because of the ambiguity between the spreading and recovery rate in the exponential onset phase (see description of simple model), we chose a narrow log-normal prior for the recovery rate $\mu \sim \rm{LogNormal}(\log(1/8), 0.2)$ with median recovery time of 8 days~\cite{maier2020}.
Note that our implementation of $\mu$ accounts for the recovery of infected people and isolation measures because it describes the duration during which a person can infect others. For the spreading rate, we assume a broad log-normal prior distribution $\lambda \sim \rm{LogNormal}(\log(0.4), 0.5)$ with median $0.4$. This way, the prior for $\lambda-\mu$ has median $0.275$ and the prior for the base reproduction number ($R_0=\lambda/\mu$) has median $3.2$, consistent with the broad range of previous estimates~\cite{zhang2020a,liu2020a,kucharski2020}. In addition, we chose a log-normal prior for the reporting delay $D\sim \rm{LogNormal}(\log(8), 0.2)$ to incorporate both the incubation time between 1--14 days with median 5~\cite{lauer2020} plus a delay from infected people waiting to contact the doctor and get tested.

The remaining model parameters are constrained by uninformative priors, in practice the Half-Cauchy distribution~\cite{gelman2006}. The half-Cauchy distribution $\rm{HalfCauchy}(x,\beta)=2/\pi\beta[1+(x/\beta)^2]$ is essentially a flat prior from zero to $O(\beta)$ with heavy tails beyond. Thereby, $\beta$ merely sets the order of magnitude that should not be exceeded for a given parameter. We chose for the number of initially infected people in the model (16 days before first data point) $I_0\sim\rm{HalfCauchy}(100)$ assuming an order of magnitude $O(100)$ and below. In addition, we chose the scale factor of the width of the likelihood function as $\sigma \sim \rm{HalfCauchy}(10)$; this choice means that the variance in reported numbers may be up to a factor of 100 larger than the actual reported number.

\paragraph{Priors for the full model (Table~\ref{tab:prior_full}):} 

\begin{table}[h]
    \centering
    \caption{Priors on the model parameters for the SIR model with change points and weekly reporting modulation.}
    \begin{tabular}{r|c|l}
        Parameter & Variable & Prior distribution \\
        \hline
        Change points & $t_1$ & $\rm{Normal}(2020/03/09,3)$\\
        \phantom{change points} & $t_2$ & $\rm{Normal}(2020/03/16,1)$\\
        \phantom{change points} & $t_3$ & $\rm{Normal}(2020/03/23,1)$\\
        Change duration & $\Delta t_i$ & $\rm{LogNormal}(\log(3),0.3)$\\
        Spreading rates & $\lambda_0$ & $\rm{LogNormal}(\log(0.4), 0.5)$\\  
        \phantom{spreading rates} & $\lambda_1$ & $\rm{LogNormal}(\log(0.2), 0.5)$\\
        \phantom{spreading rates} & $\lambda_2$ & $\rm{LogNormal}(\log(1/8), 0.5)$\\
        \phantom{spreading rates} & $\lambda_3$ & $\rm{LogNormal}(\log(1/16), 0.5)$\\
        Recovery rate & $\mu$ & $\rm{LogNormal}(\log(1/8), 0.2)$\\
        Reporting delay & $D$ & $\rm{LogNormal}(\log(8), 0.2)$\\
        Weekly modulation amplitude & $f_w$ & $\rm{Beta}(\rm{mean} = 0.7, \rm{std}=0.17)$\\
        Weekly modulation phase & $\Phi_w$ & $\rm{vonMises}(\rm{mean}=0, \kappa=0.01)$ (nearly flat)\\
        \hline
        Initially infected & $I_0$ & $\rm{HalfCauchy}(100)$\\
        Scale factor & $\sigma$ & $\rm{HalfCauchy}(10)$\\
    \end{tabular}
    \label{tab:prior_full}
\end{table}

In order to constrain our full model, an SIR model with weekly reporting modulation and  change points in the spreading rate, we chose the same priors as for the simple model but added the required priors associated with the change points. In general, we assume that each set of governmental interventions (together with the increasing awareness) leads to a reduction (and not an increase) of $\lambda$ at a change point. As we cannot know yet the precise reduction factor, we adhere to assume a reduction by $\approx 50 \%$, but always with a fairly wide 
uncertainty, so that in principle even an increase at the change point would be possible. We model the time dependence of $\lambda$ as change points, and not as continuous changes because the policy changes were implemented in three discrete steps, which were presumably followed by the public in a timely fashion.

For the spreading rates, we chose log-normal distributed priors as in the simple model. In particular, we chose for the initial spreading rate the same prior as in the simple model, $\lambda_0 \sim \rm{LogNormal}(\log(0.4), 0.5)$; after the first change point $\lambda_1 \sim \rm{LogNormal}(\log(0.2), 0.5)$, assuming the first intervention to reduce the spreading rate by $50\%$ from our initial estimates ($\lambda_0\approx0.4$) with a broad prior distribution; after the second change point $\lambda_2 \sim \rm{LogNormal}(\log(1/8), 0.5)$, assuming the second intervention to reduce the spreading rate to the level of the recovery rate, which would lead to a stationary number of new infections. This corresponds approximately to a reduction of $\lambda$ at the change point by $50\%$; and after the third change point $\lambda_3 \sim \rm{LogNormal}(\log(1/16), 0.5)$, assuming the third intervention to reduce the spreading rate again by $50\%$. With that intervention, $\lambda_3$ is smaller than the recovery rate $\mu$, causing a decrease in new case numbers and a saturation of the cumulative number of infections.

For the timing of change points, we chose normally distributed priors. In particular, we chose $t_1\sim \rm{Normal}(2020/03/09,3)$ for the first change point because on the weekend of March 8, large public events, like soccer matches or fairs, were  cancelled. For the second change point, we chose $t_2\sim \rm{Normal}(2020/03/16,1)$, because on March 15,  the closing  of schools and other educational institutions along with the closing of non-essential stores were announced and implemented on the following day. Restaurants were allowed to stay open until 6 pm. For the third change point, we chose $t_3\sim \rm{Normal}(2020/03/23,1)$, because on March 23, a far-reaching contact ban (“Kontaktsperre”), which includes the prohibition of even small public gatherings as well as complete closing of restaurants and non-essential stores was imposed by the government authorities. Further policy changes may occur in future; however, for now, we do not include more change points.

The change points take effect over a certain time period $\Delta t_i$ for which we choose $\Delta t_i\sim \rm{LogNormal}(\log(3),0.3)$ with a median of 3 days over which the spreading rate changes continuously as interventions have to become effective. The precise duration of the transition has hardly any affect on the cumulative number of cases (Fig.~\ref{fig:timing}~E-F). We assumed a duration of three days, because some policies were not announced at the same day for all states within Germany; moreover, the smooth transition also can absorb continuous changes in behavior.

The number of tests that are performed and reported vary regularly over the course of a week and are especially low during weekends. To account for this periodic variation, we modulated the number of inferred cases by the absolute value of a sine function with in total a period of 7 days. We chose this function as it is a non-symmetric oscillation, fitting the weekly variation of cases on a phenomenological level. For the amplitude of the modulation we chose a weakly informative Beta prior: $f_w \sim \rm{Beta}(\rm{mean} = 0.7, \rm{std}=0.17 )$ and for the phase a nearly flat circular distribution: $\Phi_w \sim  \rm{vonMises}(\rm{mean}=0, \kappa=0.01)$. 

\subsection*{Model comparison}
Since change point detection entails evaluating models with different numbers of parameters, some form of fair model comparison is needed. This is necessary to compensate for the higher flexibility of more complex models, as this flexibility carries the risk of overfitting and overconfident forecasts. The standard approach to avoid over-fitting in machine learning is cross-validation, and cross validation has recently also been advocated for Bayesian model comparison (e.g. ~\cite{vehtari2017,gelman2013}), especially for models employed for predictions and forecasts. Thus, one would ideally like to compare the models with different numbers of change points by the probability they assign to previously unobserved data points. Technically this is measured by their out-of-sample prediction accuracy, i.e. their log pointwise predictive density (lppd):
\begin{equation}
    \mathrm{lppd} = \sum_{i=1}^N \log \left( \int p(y_i^{\mathrm{os}}|\theta)p_{\mathrm{post}}(\theta) d\theta \right)    ~,
    \label{eq:lppd}
\end{equation}
\noindent where the vector $[y_1^{\mathrm{os}},\ldots,y_1^{\mathrm{os}}]$ is a an out-of-sample dataset of $N$ new data points, and where $p_{\mathrm{post}}(\theta)=p_{\mathrm{post}}(\theta|y,M_j)$ is the posterior distribution of the parameters, given the in-sample data $y$ and the model $M_j$. In practice, the integral is approximated using a sufficient amount of samples from $p_{\mathrm{post}}(\theta)$. However, this approach is only reasonable if a sufficient amount of out-of-sample data is available, which is not the case in the early stages of a disease outbreak. Therefore, the pointwise out-of-sample prediction accuracy was approximated using Leave-one-out cross-validation (LOO) in PyMC3 to compute equation \ref{eq:lppd} individually for each left out data point based on the model fit to the other data points. The sum of these values, multiplied by a factor of $-2$ then yields the leave-one-out cross-validation (LOO-CV) score. Thus, lower LOO-CV scores imply better models.

\bibliography{covid}

\bibliographystyle{Science}

\section*{Acknowledgments}
We thank Tim Friede, Theo Geisel, Knut Heidemann, Moritz Linkmann, Matthias Loidolt, and Vladimir Zykov for carefully and promptly reviewing our work internally. We thank the Priesemann group - especially Daniel Gonzalez Marx, Fabian Mikulasch, Lucas Rudelt \& Andreas Schneider - for exciting discussions and for their valuable comments. We thank the colleagues of the G\"ottingen Campus, with whom we were discussing the project and the COVID-19 case forecast in the past weeks very intensively: Heike Bickeb\"oller, Eberhard Bodenschatz, Wolfgang Br\"uck, Alexander Ecker, Andreas Leha,  Ramin Golestanian, Helmut Grubm\"uller, Stephan Herminghaus, Reinhard Jahn, Norbert Lossau \& Simone Scheithauer. We thank Nils Bertschinger for stimulating discussion on Bayesian modeling and model comparison.
\textbf{Funding:} All authors received support from the Max-Planck-Society. JD and PS acknowledge funding by SMARTSTART, the joint training program in computational neuroscience by the VolkswagenStiftung and the Bernstein Network. JZ received financial support from the Joachim Herz Stiftung. MW is employed at the Campus Institute for Dynamics of Biological Networks funded by the VolkswagenStiftung.
\textbf{Author contributions:} JD, JZ, MW, MW, VP designed research. JD, JZ, PS conducted research. JD, JZ, PS, JPN, MW, MW, VP analyzed the data. JD, PS, JPN created figures. All authors wrote the paper.  
\textbf{Competing interests:} The authors declare no competing interests.
\textbf{Data and materials availability:} Both data~\cite{data2020} and analysis code~\cite{dehning2020b} are available online. This work is licensed under a Creative Commons Attribution 4.0 International (CC BY 4.0) license, which permits unrestricted use, distribution, and reproduction in any medium, provided the original work is properly cited. To view a copy of this license, visit https://creativecommons.org/licenses/by/4.0/. This license does not apply to figures/photos/artwork or other content included in the article that is credited to a third party; obtain authorization from the rights holder before using such material.


\section*{List of Supplementary materials:}
\noindent
Table S1 -- S2\\
Fig S1 -- S7\\

\clearpage
\section*{Supplementary Materials:}
\renewcommand{\thefigure}{S\arabic{figure}}
\setcounter{figure}{0}
\renewcommand{\thetable}{S\arabic{table}}
\setcounter{table}{0}


\begin{table}[h]
    \centering
    \caption{Overview of model parameters.}
    \begin{tabular}{l|l}
        Variable & Parameter \\
        \hline
        $\lambda$                      & Spreading rate \\
        $\mu$                          & Recovery rate \\
        $\lambda^* = \lambda-\mu$      & Effective spreading rate \\
        $\lambda_i$                    & Spreading rate after $i$-th  intervention\\
        $t_i$                          & Time of $i$-th intervention \\
        $f_w$                          & Amplitude of weekend corrections\\
        $\Phi_w$                       & Phase shift of weekend correction\\
        $\sigma$                       & Scale factor of the width of Student's t-distribution \\
        \hline
        $N$                            & Population size (83.000.000) \\
        $S_t$                          & Susceptible at time $t$ \\
        $I_t$                          & Infected at time $t$ \\
        $R_t$                          & Recovered at time $t$ \\
        $\Delta t$                     & Time step \\
        $R_t^\mathrm{new} = \mu \Delta t I_{t-1}$  & New recoveries at time $t$ \\
        $I_t^\mathrm{new} = \lambda \Delta t\frac{S_{t-1}}{N} I_{t-1}$ & New infections at time $t$ \\
        $C_t = I_{t-D}^\mathrm{new}$       & New reported cases at time $t$ \\
        $D$                                & Delay of case detection \\
    \end{tabular}
    \label{tab:parameters}
\end{table}

\begin{table}[h!]
    \centering
    \caption{Model comparison: Using leave-one-out (LOO) cross-validation, we compare our original ``SIR main'' model that features a weekend modulation (to account for fewer reported cases during weekends) with other model variants. Remarkably, the median inferred effective growth rate $\lambda^\ast$ after the last change-point is very similar for all model variants.
    For full details on the model variants, see the figure captions in the SI:
    (i) SEIR-like with explicit incubation time, Fig.~\ref{fig:inference_SEIR_not_rw}.
    (ii) SIR excluding the weekend modulation, Fig.~\ref{fig:SIR_without_modulation}.
    (iii) Sensitivity analysis by applying wider priors to different parameters, Figs.~\ref{fig:SIR_wider_delay},~\ref{fig:SIR_wider_days},~\ref{fig:SIR_wider_transient}.
    Lower LOO-scores represent a better match between model and data (pLOO is the effective number of parameters).  $^\dagger$For the SEIR-like model, the magnitude of the effective growth rate is not directly comparable because of the explicit incubation period.}
    \begin{tabular}{l|c||c|c|c|c||l|l}
        Model & \# c-pts. &
        $\lambda^\ast_0$ & $\lambda^\ast_1$ & $\lambda^\ast_2$ & $\lambda^\ast_3$ &
        LOO-score & pLOO \\
        \hline
        SIR main & 0 &
        0.03 &&&&
        $927 \pm 9$  & $8.31$\\

        SIR main & 1 &
        0.21 & -0.03 &&&
        $819 \pm 16$  & $13.46$\\

        SIR main & 2 &
        0.30 & 0.11 & -0.03 &&
        $796 \pm 17$  & $12.53$\\

        SIR main & 3 &
        0.30 & 0.12 & 0.02 & -0.03 &
        $787 \pm 17$  & $13.42$\\

        SIR without weekend modulation & 3 &
        0.31 & 0.13 & 0.04 & -0.03 &
        $807 \pm 17$  & $9.65$\\

        SIR with wider delay prior & 3 &
        0.30&0.12&0.02&-0.04&
        $787 \pm 17$ & $14.15$ \\

        SIR with wider change point priors & 3 &
        0.30&0.12&0.02&-0.04&
        $787 \pm 17$ & $14.29$ \\

        SIR with wider transient length prior & 3 &
        0.29&0.12&0.02&-0.04&
        $785 \pm 17$ & $13.92$ \\

        SEIR-like$^\dagger$ & 3 &
        1.95 & 0.45 & 0.05 & -0.12&
        $782 \pm 17$ & $10.71$\\ 
    \end{tabular}
    \label{tab:loo_suppls}
\end{table}


\begin{figure}[b]
    \centering
    \centerline{
    \includegraphics[scale=.8]{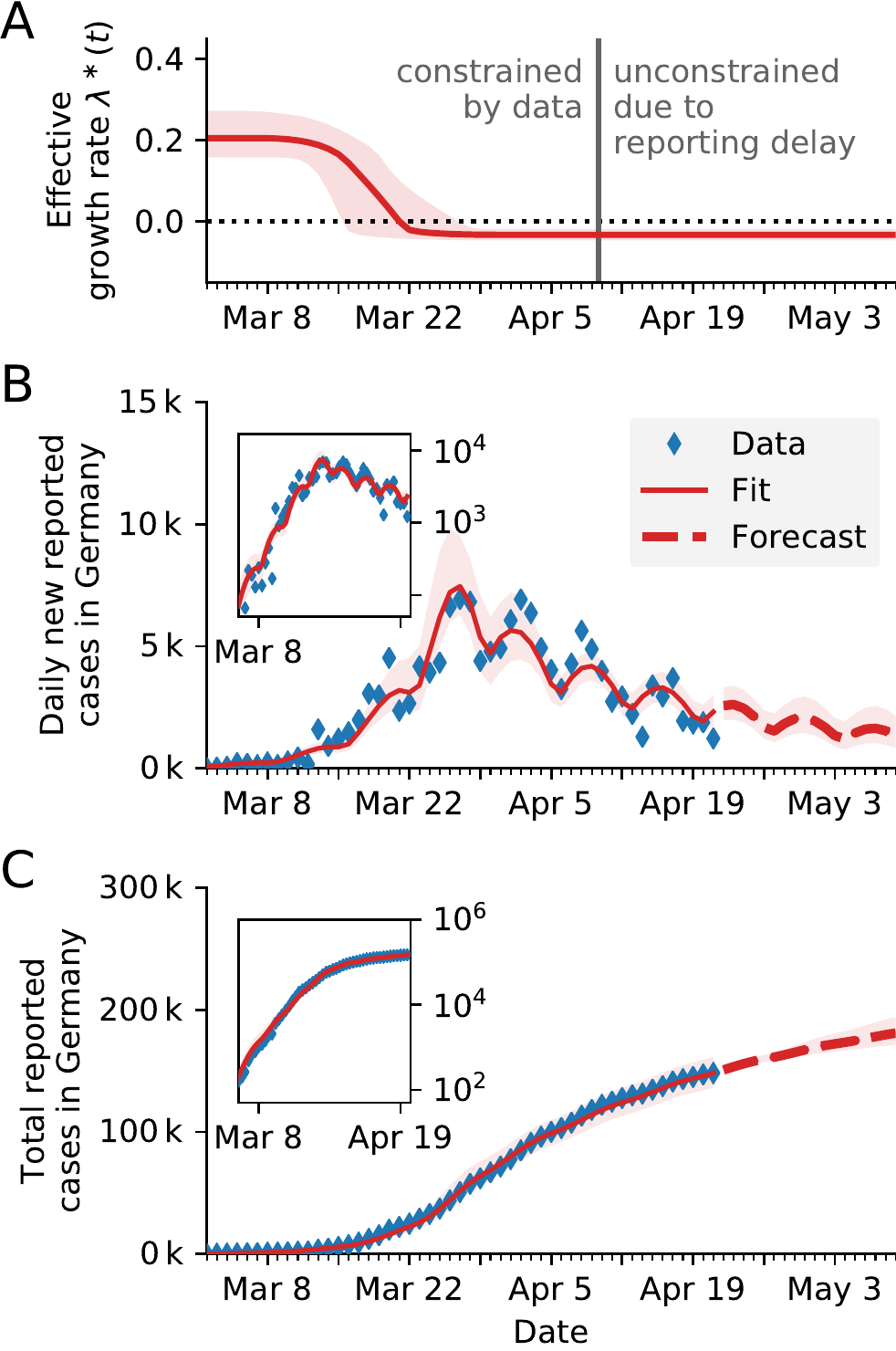}
    \hspace{2mm}
    \includegraphics[scale=.8]{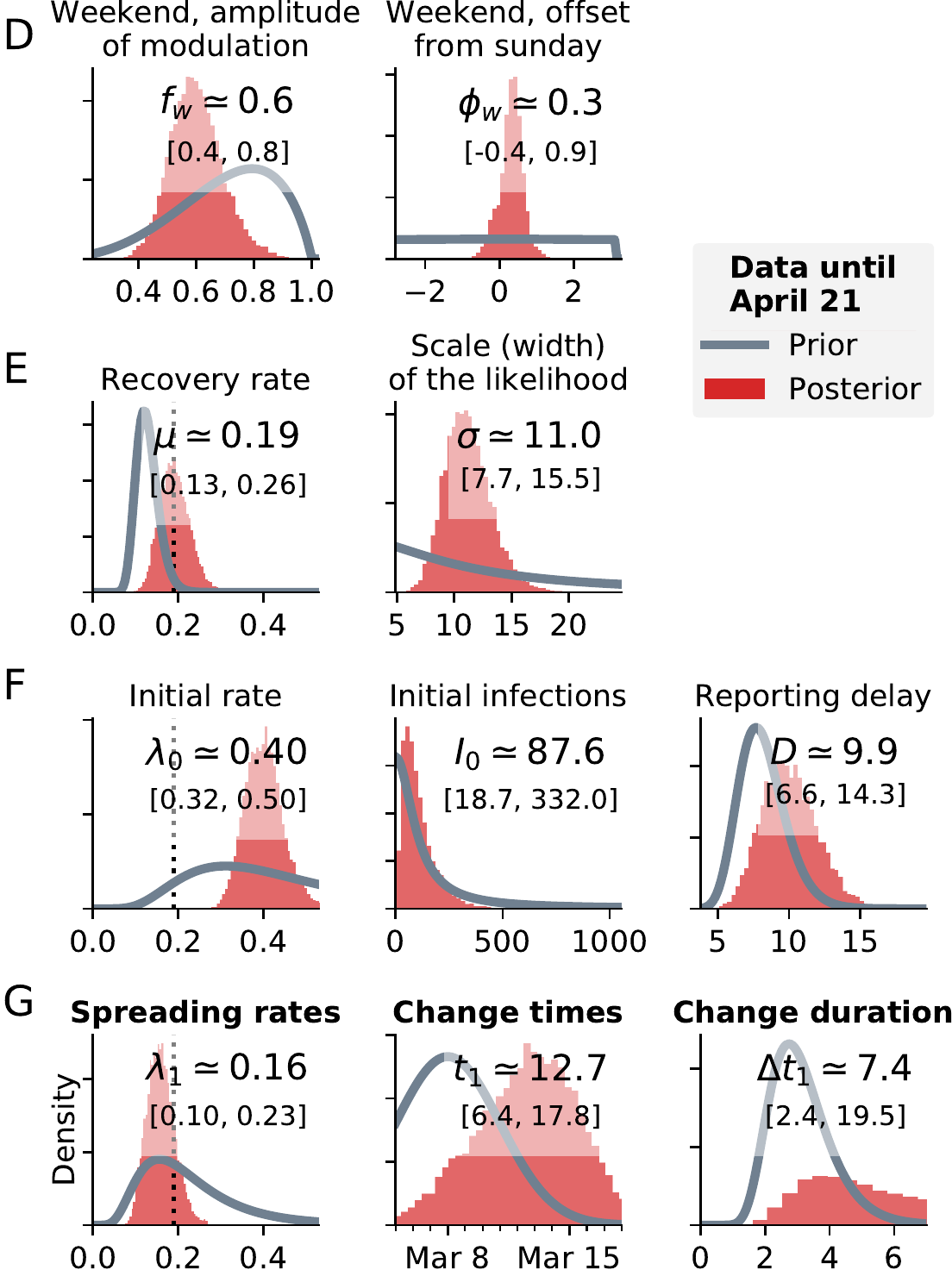}
    }
    \caption{\textbf{Model comparison:} Change-point detection as in Fig.~\ref{fig:inference_3_change_points} (three change points,  main text) with the \textbf{same model} ``SIR main'' but only \textbf{one change point}.  With only one change point, the model cannot describe the data well after April 1. \textbf{A:}~Time-dependent model estimate of the effective growth rate $\lambda^*(t)$. \textbf{B:} Forecast and comparison of daily new reported cases with the model fit, \textbf{inset:} Same on log-lin scale.
    \textbf{C:}~Same as B but for cumulative (total) cases.
    \textbf{D--G:}~Posterior distributions from the change point detection (red) compared to prior distributions (gray). Please refer to Fig.~\ref{fig:simple} (main text) for a more detailed description of the distributions.}
    \label{fig:inference_1_change_point}
\end{figure}

\begin{figure}[b]
    \centering
    \centerline{
    \includegraphics[scale=.8]{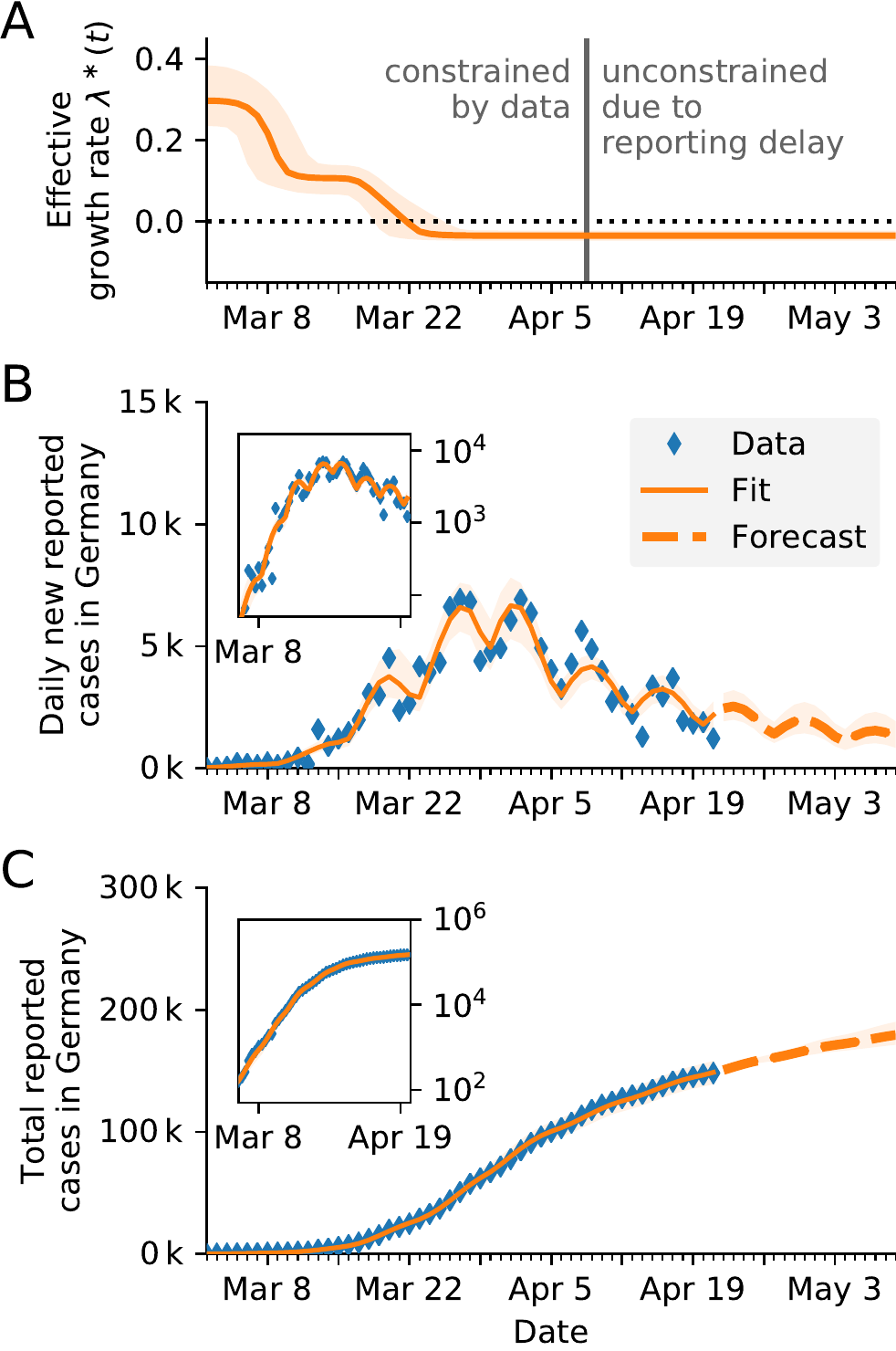}
    \hspace{2mm}
    \includegraphics[scale=.8]{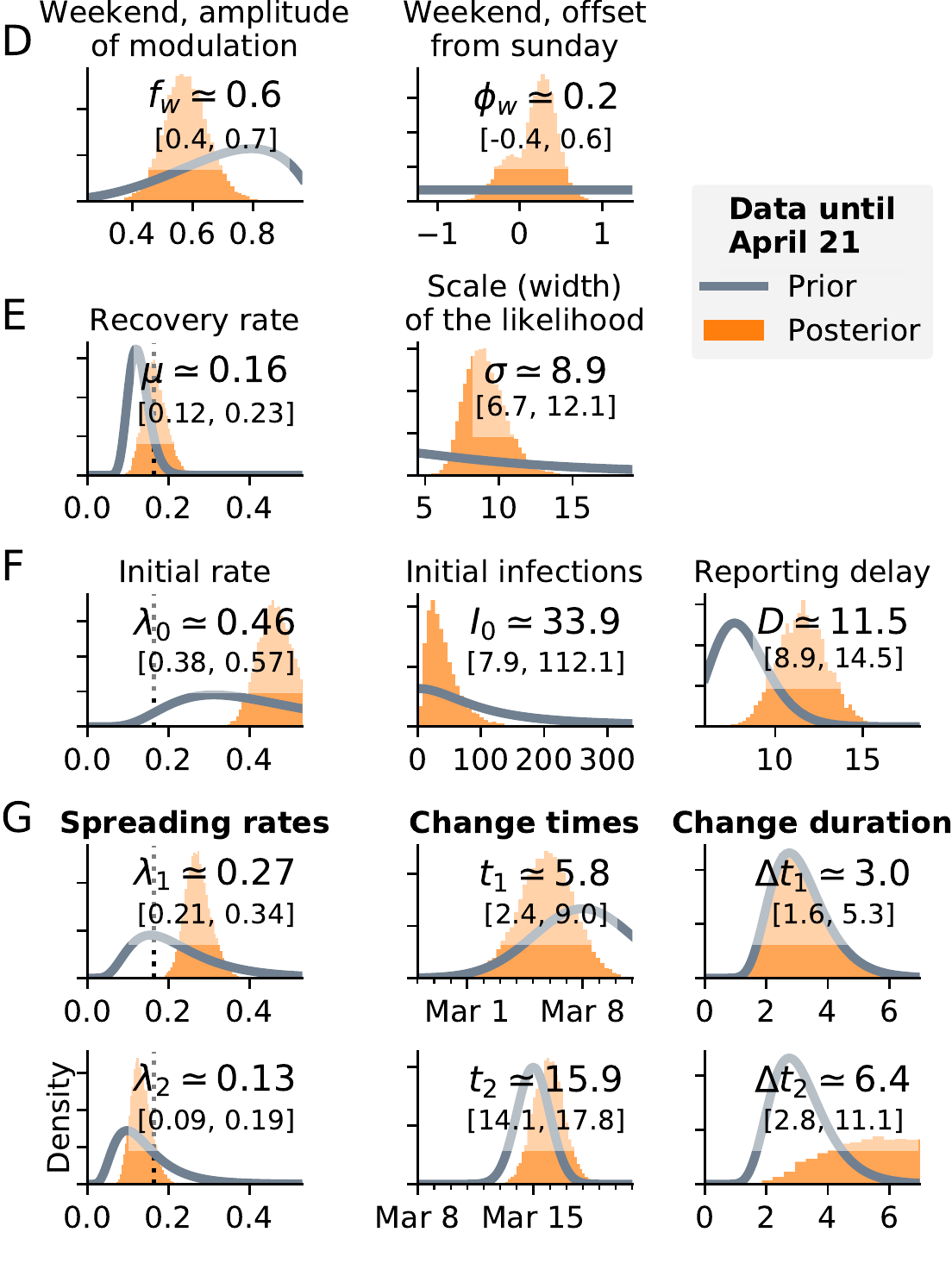}
    }
    \caption{\textbf{Model comparison:} Change-point detection as in Fig.~\ref{fig:inference_3_change_points} (three change points,  main text) with the \textbf{same model} ``SIR main'' but with \textbf{two change points}.
    With two change points, the onset time of the second change point is close to the (middle) one inferred when using the 3-c.p.~model. However, the effective growth rate $\lambda^\ast$ after the respective last c.p.~is the same in both cases, $\lambda^\ast=-0.03$ (see also Table~\ref{tab:loo_suppls}).
    \textbf{A:} Time-dependent model estimate of the effective growth rate $\lambda^*(t)$. Two change points are clearly visible. \textbf{B:} Forecast and comparison of daily new reported cases with the model fit, \textbf{inset:} Same on log-lin scale.
    \textbf{C:}~Same as B but for cumulative (total) cases.
    \textbf{D--G:}~Posterior distributions from the change point detection (orange) compared to prior distributions (gray). Please refer to Fig.~\ref{fig:simple} (main text) for a more detailed description of the distributions.}
    \label{fig:inference_2_change_points}
\end{figure}

\begin{figure}[b]
    \centering
        \centerline{
    \includegraphics[scale=.7]{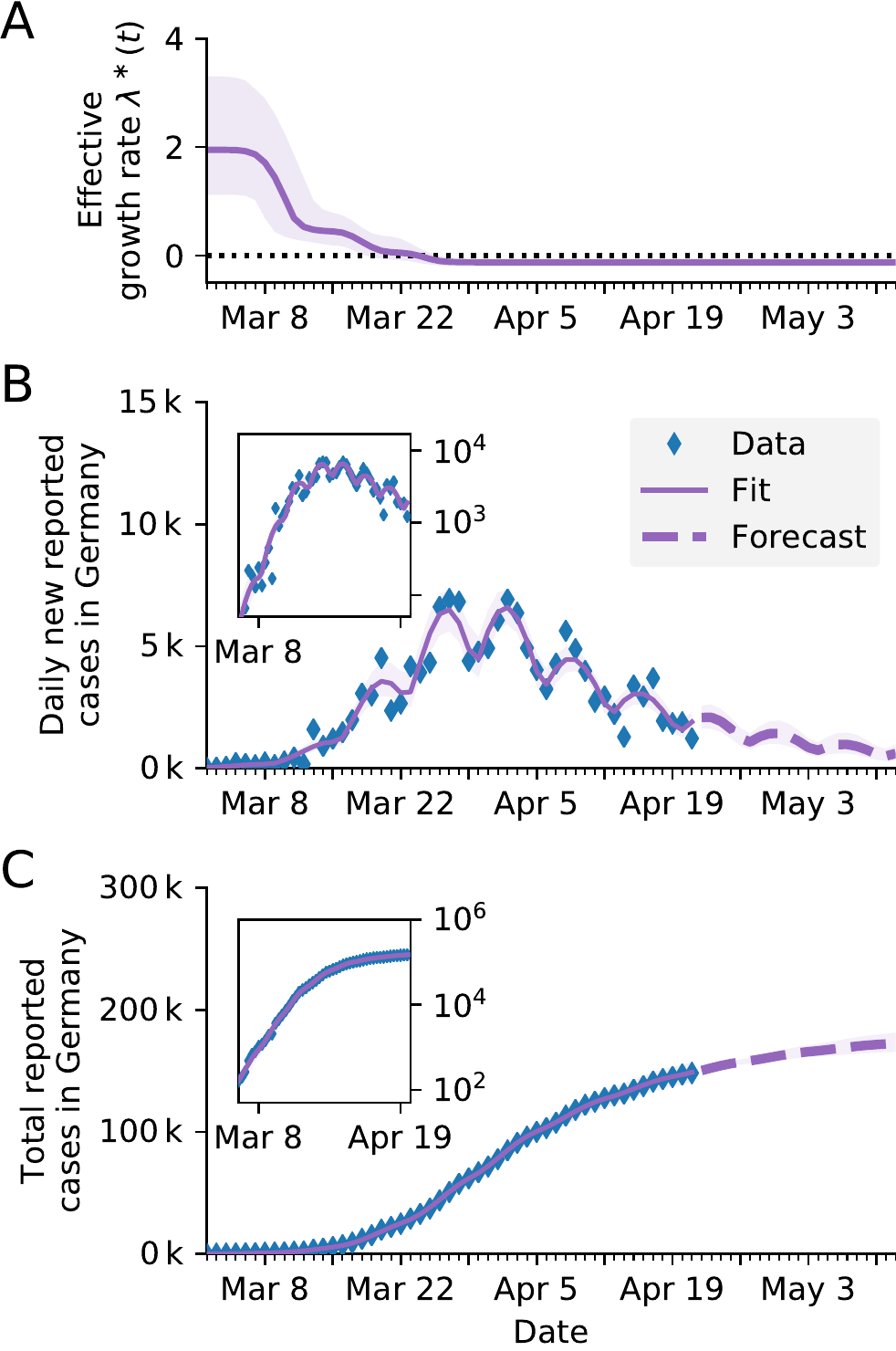}
    \hspace{2mm}
    \includegraphics[scale=.7]{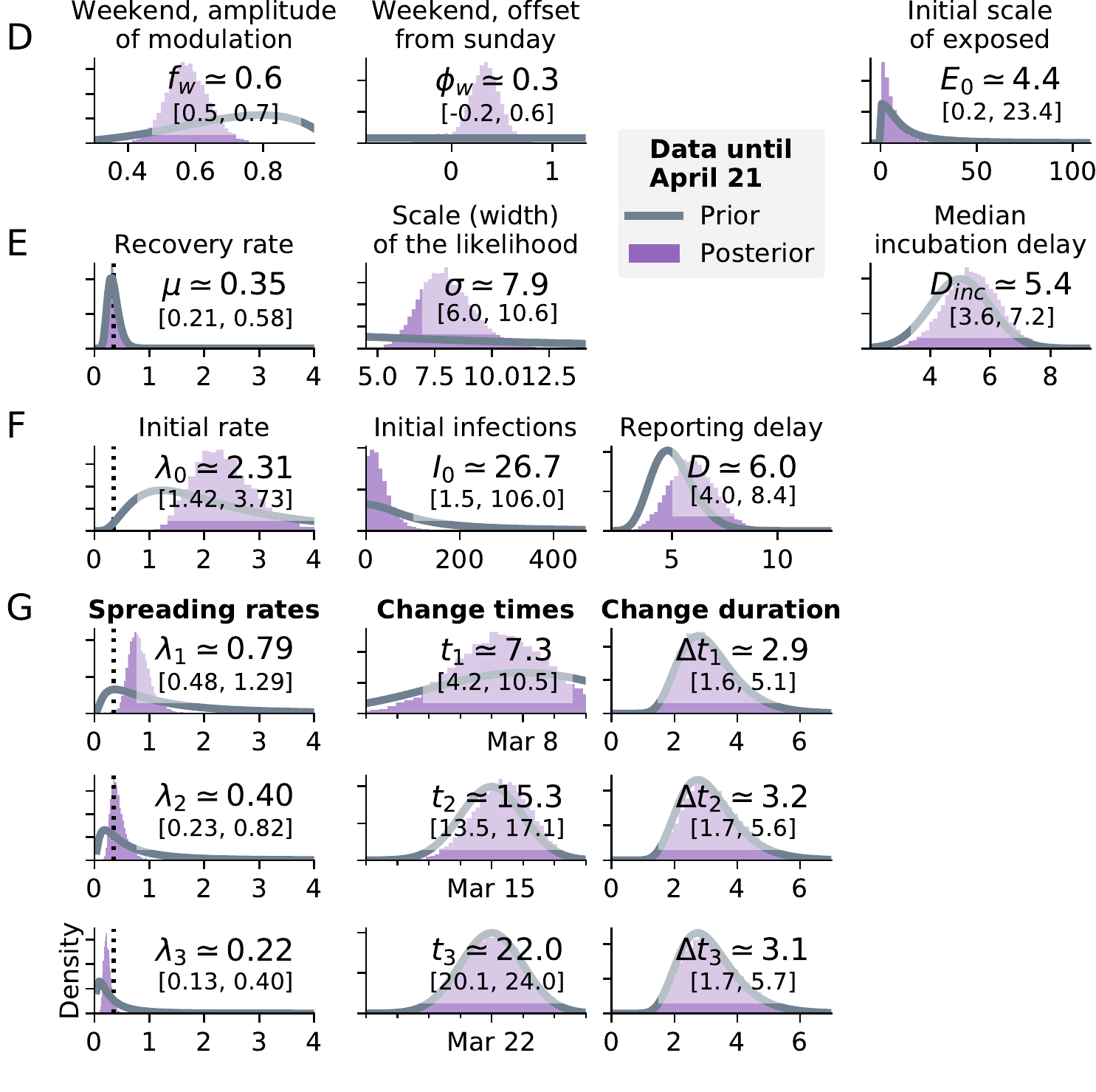}
    }
    \caption{
    \textbf{Model comparison:} Change-point detection as in Fig.~\ref{fig:inference_3_change_points} (three change points,  main text) but with a more involved \textbf{SEIR-\textit{like}} model and \textbf{three change points}, code available online~\cite{dehning2020b}.
    Arguably, the SEIR-like model provides are more realistic (but also more complex) description of virus propagation. It yields a slightly better (lower) LOO-score than our ``SIR main'' model in the cross-validation, Table~\ref{tab:loo_suppls}. However, inferred parameters are compatible with the simpler model and the inferred dynamics are similar.
    \textbf{Model details}:~The SEIR-like model builds on our ``SIR main'' model (which includes a weekend correction, see Methods). The SEIR-like model features an explicit log-normal incubation period and a lognormal reporting delay.  The incubation period is implemented as a discrete convolution of multiple exposed pools with a lognormal kernel. The discrete lognormal kernel is parameterized as follows (to match the characteristic incubation time of COVID-19~\cite{lauer2020}): median $D_\mathrm{inc}$, scale parameter 0.418 and normalized to 1. The median is a free parameter with prior $\mathrm{Normal}(5,1)$ (days)~\cite{lauer2020}. 
    The reporting delay is implemented in a similar manner: as a convolution of the number of new cases with a lognormal kernel and the scale  parameter is fixed to 0.3. In order to match the total delay of the main model (between the infection and the observation), the median $D$ is a free parameter with prior $\mathrm{LogNormal}(5,0.2)$ (days). 
    Note that because of the lognormal-distributed incubation period in the SEIR-like model, the spreading and recovery rates are not directly comparable to the SIR models. We correspondingly adapted the respective priors~\cite{li2020b}: We adapted the prior median of the recovery rate $\mu$ to 1/3 with a scale factor of 0.3, and the prior median for $\lambda_0$, $\lambda_1$, $\lambda_2,$ and $\lambda_3$ to 2.0, 1.0, 0.5 and 0.25, respectively, with a scale parameter of 1 each.
    \textbf{A:}~Time-dependent model estimate of the effective growth rate $\lambda^*(t)$. \textbf{B:}~Comparison of daily new reported cases and the model (inset: log-lin scale). \textbf{C:}~As B but for total (cumulative) reported cases. \textbf{D--G}:~Prior and posterior distributions of all free parameters.}
    \label{fig:inference_SEIR_not_rw}
\end{figure}

\begin{figure}[b]
    \centering
        \centerline{
    \includegraphics[scale=.7]{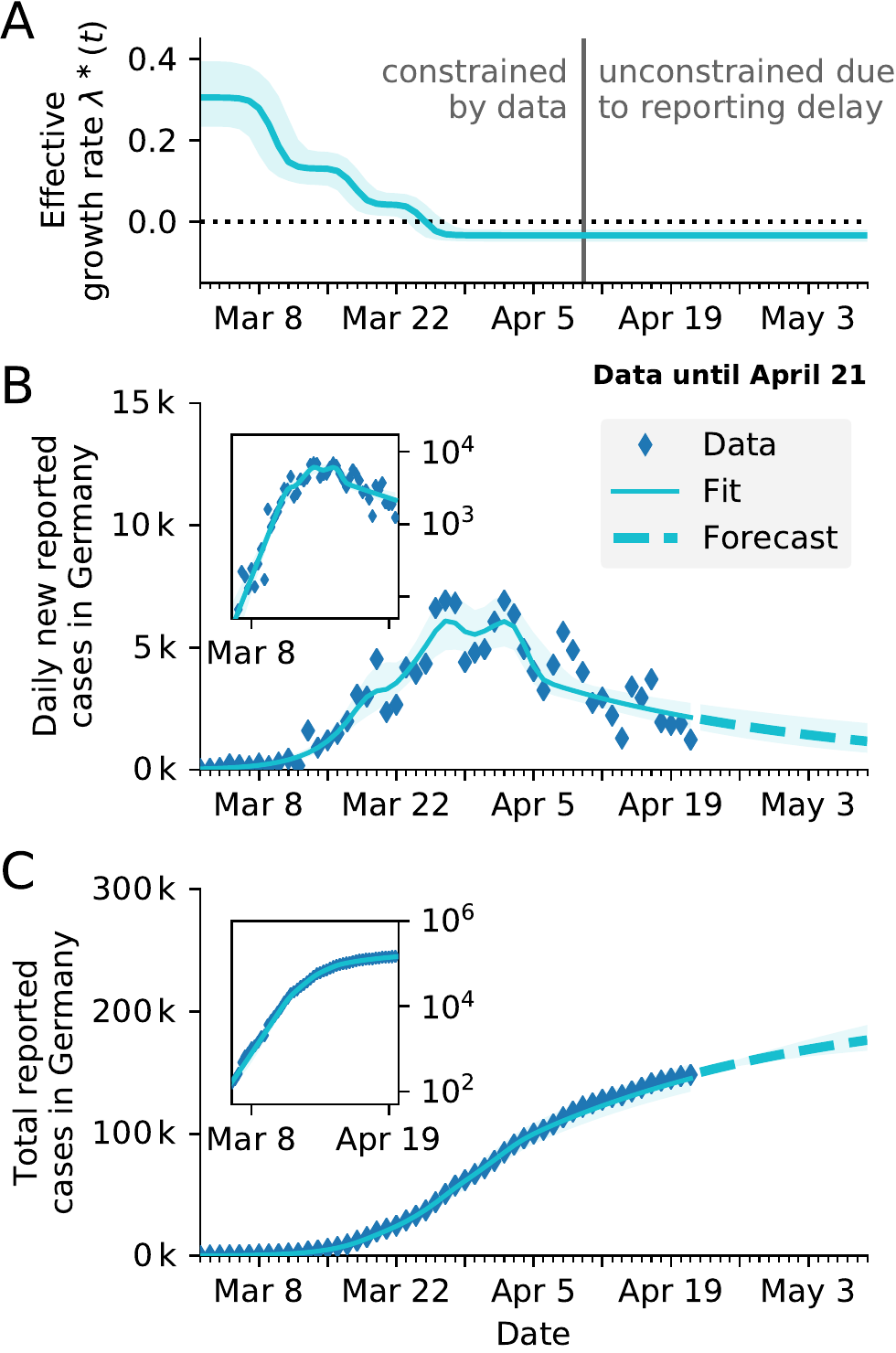}
    \hspace{2mm}
    \includegraphics[scale=.7]{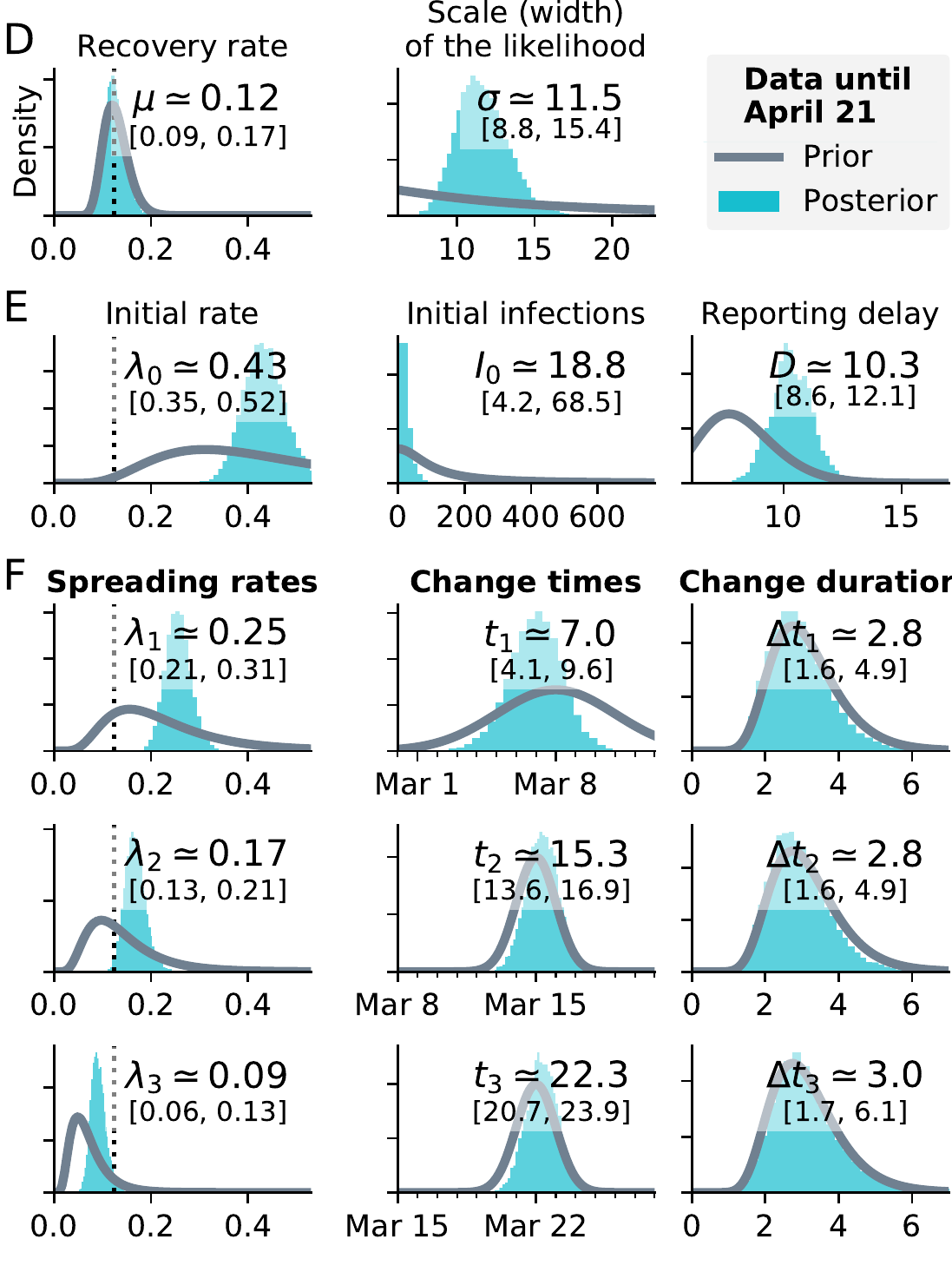}
    }
    \caption{
    \textbf{Model comparison:} Change-point detection as in Fig.~\ref{fig:inference_3_change_points} (three change points,  main text) with an \textbf{SIR model that excludes the weekend modulation} but features the \textbf{three change points}.
    In this version of the model, we excluded the assumption that daily new reported cases depend on the weekday (which is modeled as an absolute sine with an amplitude and a phase shift as inferred parameters in the main model).
    While the inferred parameters from the model that excludes the weekend modulation (in particular rates and onset times of change points) match the model that includes the modulation, the LOO-scores of the cross-validation are worse, Table~\ref{tab:loo_suppls}. Especially in panel B it becomes clear that the (empirically motivated) dependence of reported on cases on the weekday is justified. Without the modulation, the model fit does not capture the periodic changes in the data.
    \textbf{A:}~Time-dependent model estimate of the effective growth rate $\lambda^*(t)$. \textbf{B:}~Comparison of daily new reported cases and the model (inset: log-lin scale). \textbf{C:}~As B but for total (cumulative) reported cases. \textbf{D--F}:~Prior and posterior distributions of all free parameters.
    }
    \label{fig:SIR_without_modulation}
\end{figure}


\begin{figure}[b]
    \centering
        \centerline{
    \includegraphics[scale=.7]{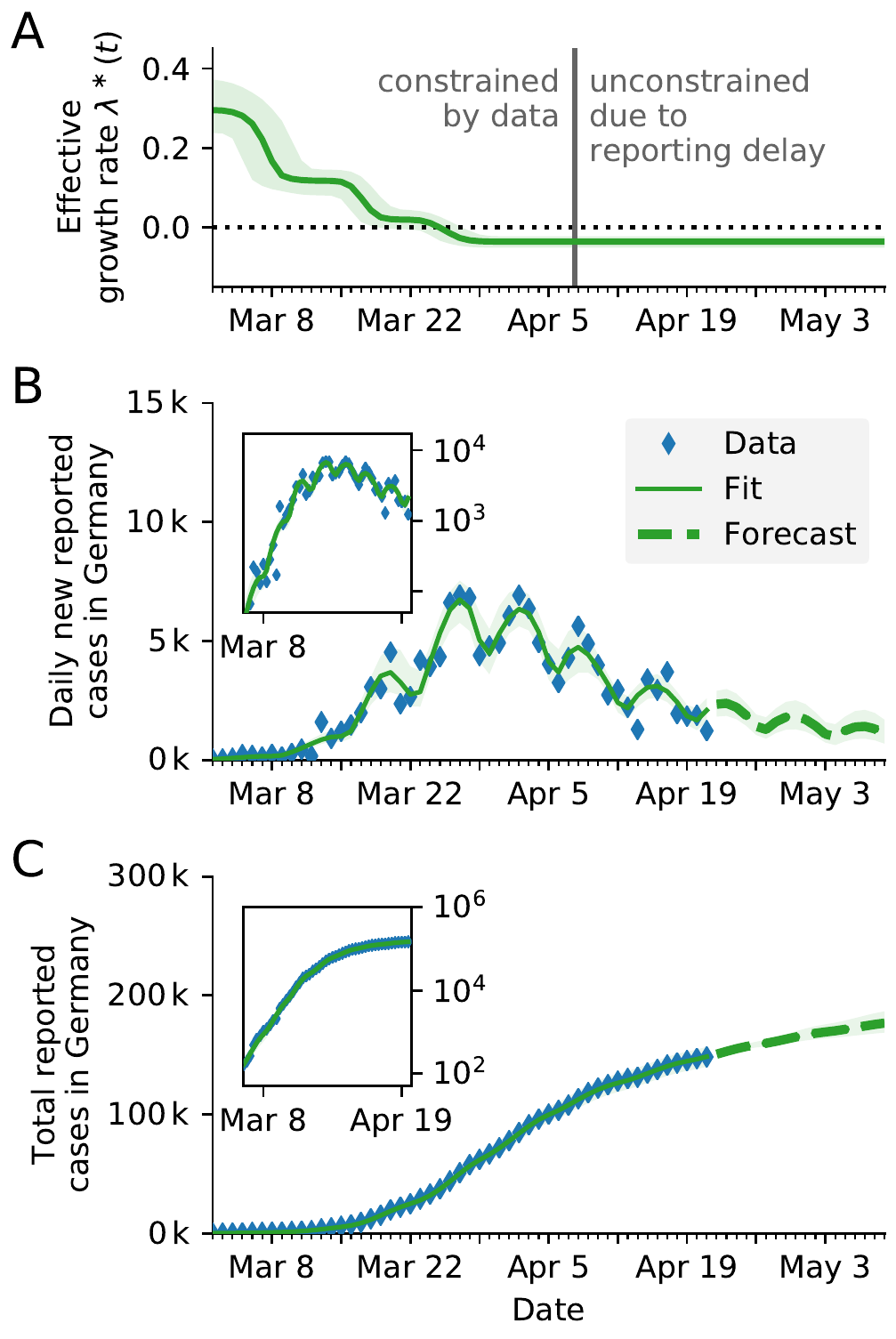}
    \hspace{2mm}
    \includegraphics[scale=.7]{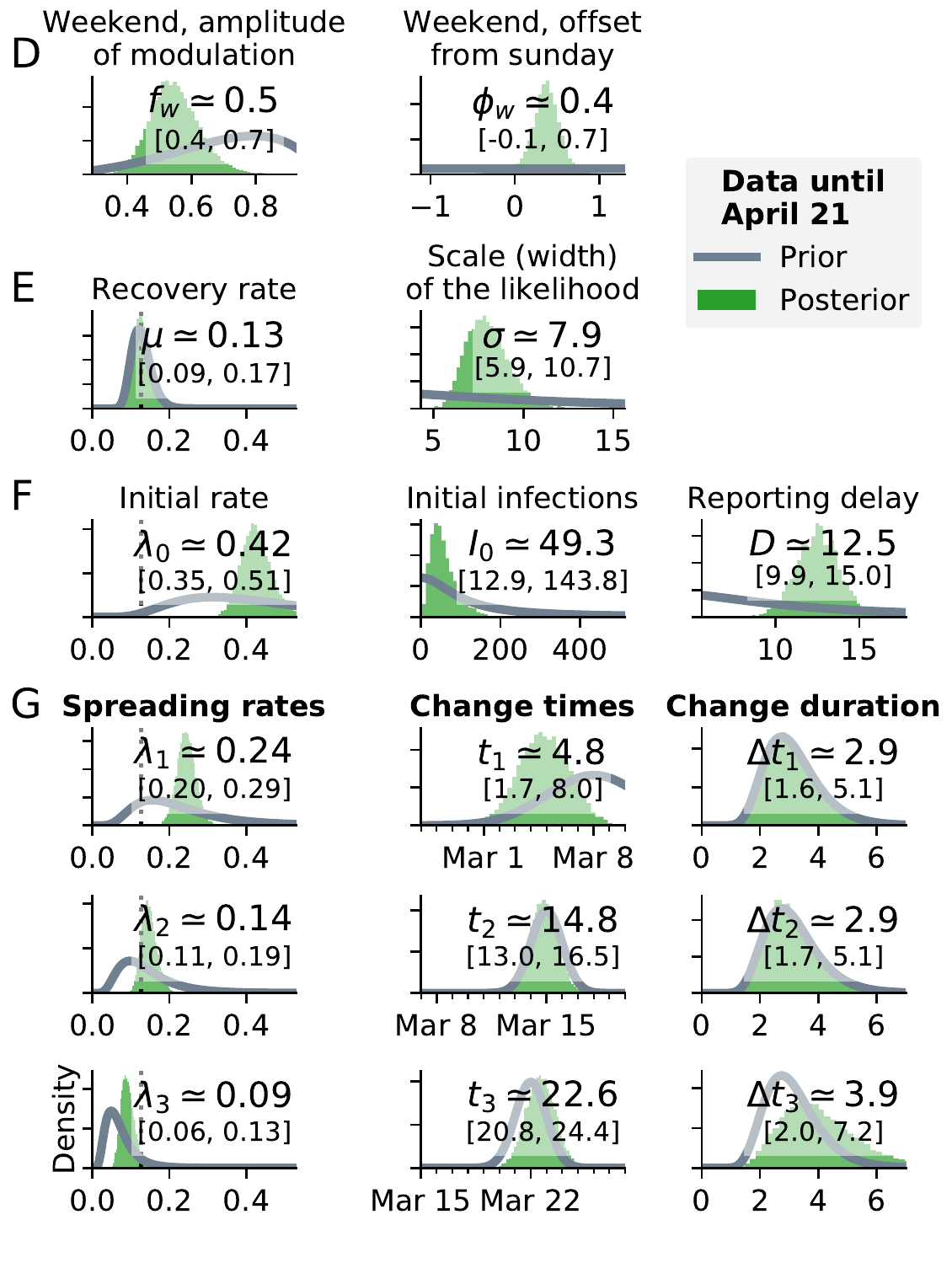}
    }
    \caption{
        \textbf{Sensitivity analysis:}
    Change-point detection as in Fig.~\ref{fig:inference_3_change_points} (main text, three change points, same model) but with \textbf{a prior for the reporting delay that is 4 times wider} (panel F, third column).
    All parameters and change points are constrained by data. In particular,  the posterior distribution of the reporting delay is slightly wider than with the original prior, but it is well constrained by data.
    \textbf{A:}~Time-dependent model estimate of the effective growth rate $\lambda^*(t)$. \textbf{B:}~Comparison of daily new reported cases and the model (inset: log-lin scale). \textbf{C:}~As B but for total (cumulative) reported cases. \textbf{D--G}:~Prior and posterior distributions of all free parameters.
    }
    \label{fig:SIR_wider_delay}
\end{figure}

\begin{figure}[b]
    \centering
        \centerline{
    \includegraphics[scale=.7]{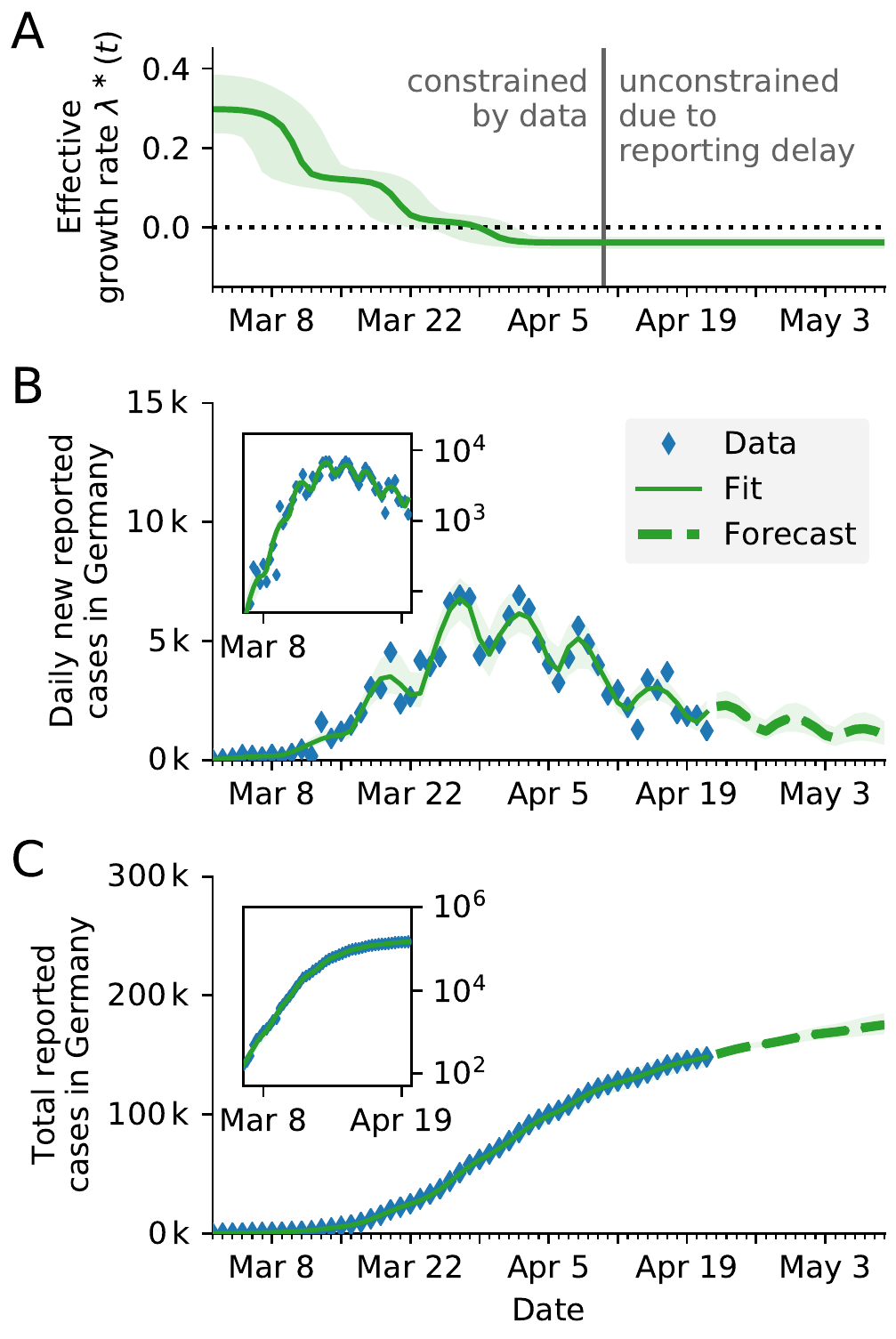}
    \hspace{2mm}
    \includegraphics[scale=.7]{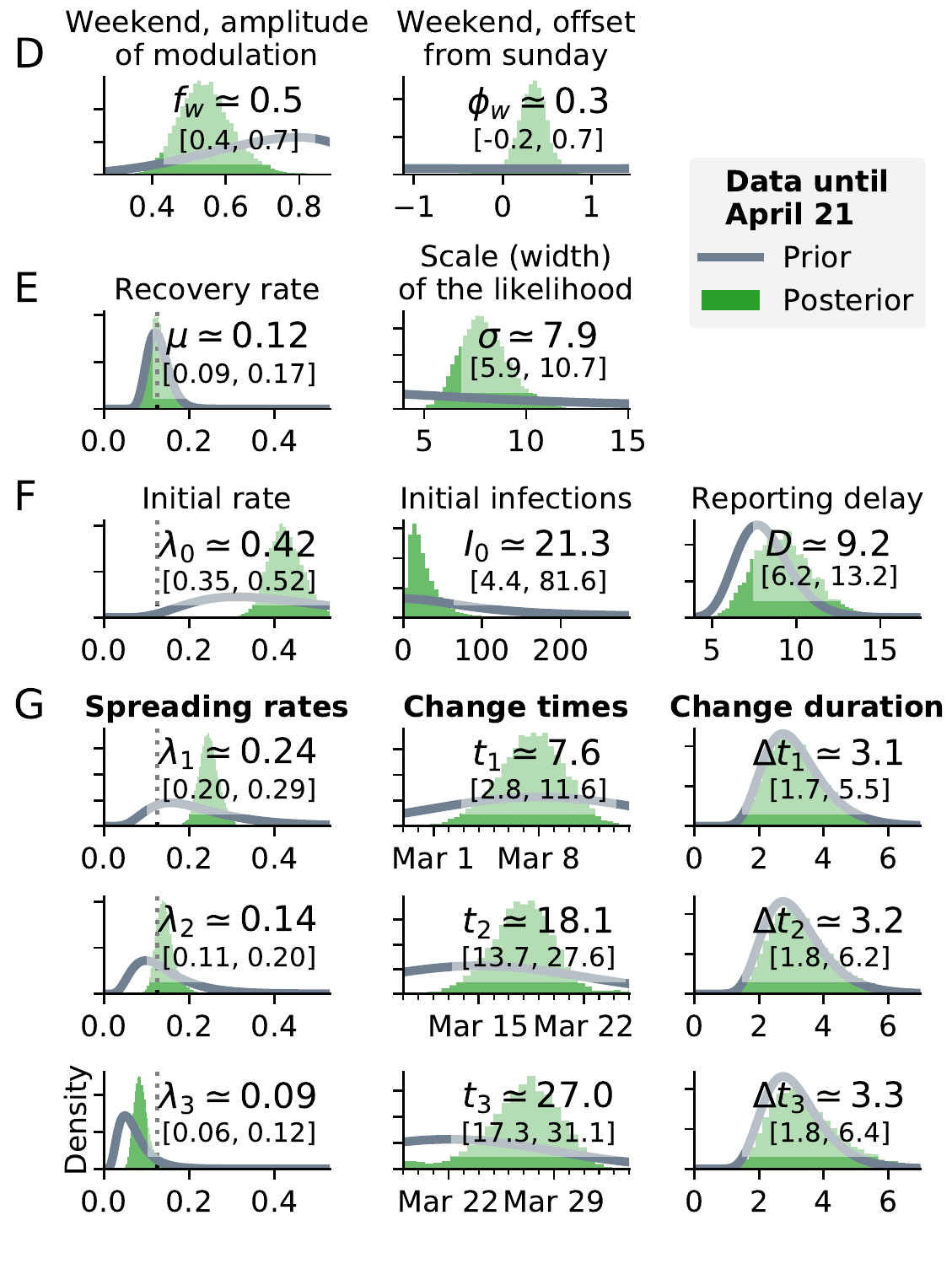}
    }
    \caption{
    \textbf{Sensitivity analysis:}
    Change-point detection as in Fig.~\ref{fig:inference_3_change_points} (main text, three change points, same model) but with \textbf{a prior for the change times that is 14 days wide}, instead of $\sim2$ days, (panel G, second column).
    All parameters and change points are constrained by data but the change points occur at later times compared to the original priors. 
    \textbf{A:}~Time-dependent model estimate of the effective growth rate $\lambda^*(t)$. \textbf{B:}~Comparison of daily new reported cases and the model (inset: log-lin scale). \textbf{C:}~As B but for total (cumulative) reported cases. \textbf{D--G}:~Prior and posterior distributions of all free parameters.
    }
    \label{fig:SIR_wider_days}
\end{figure}

\begin{figure}[b]
    \centering
        \centerline{
    \includegraphics[scale=.7]{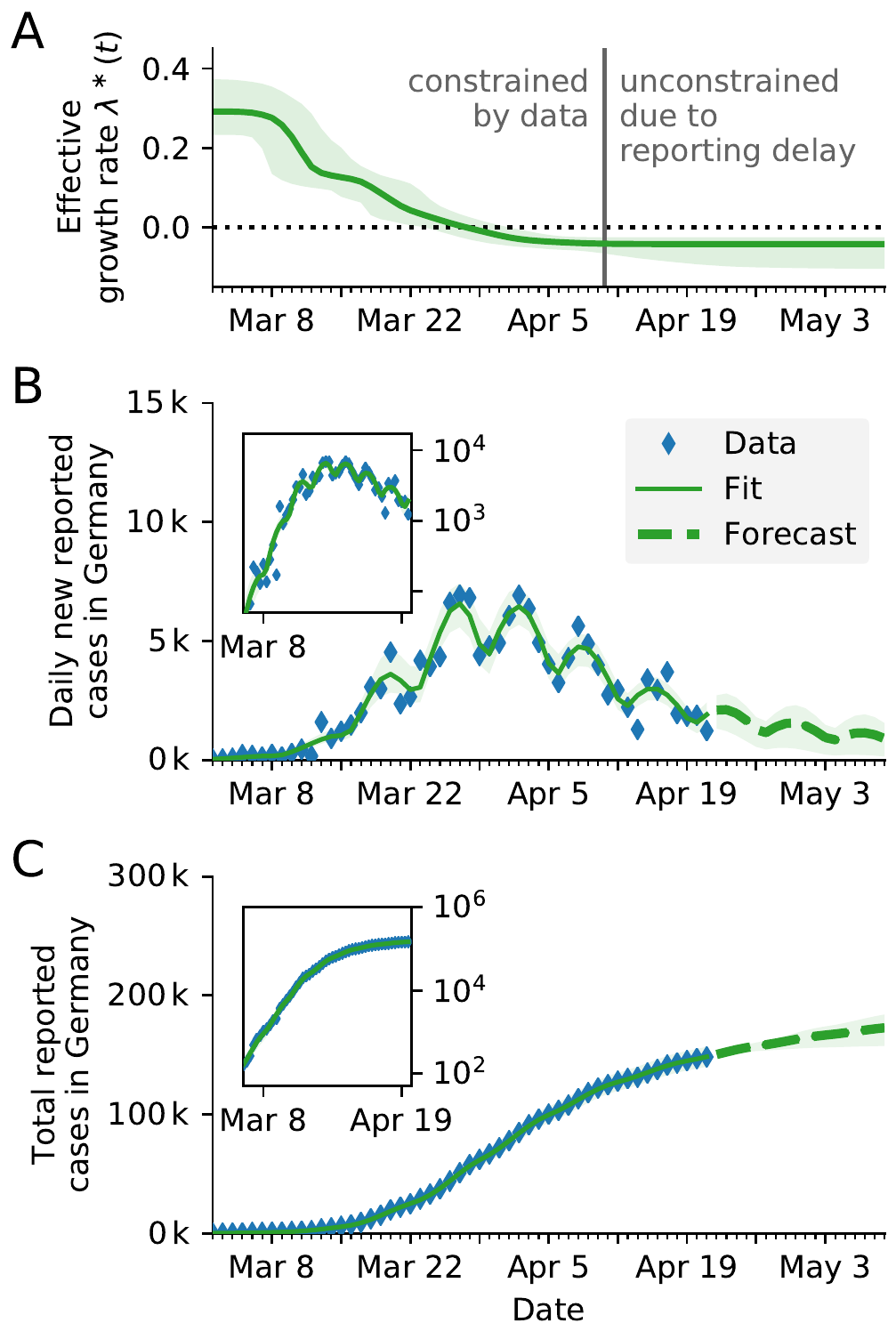}
    \hspace{2mm}
    \includegraphics[scale=.7]{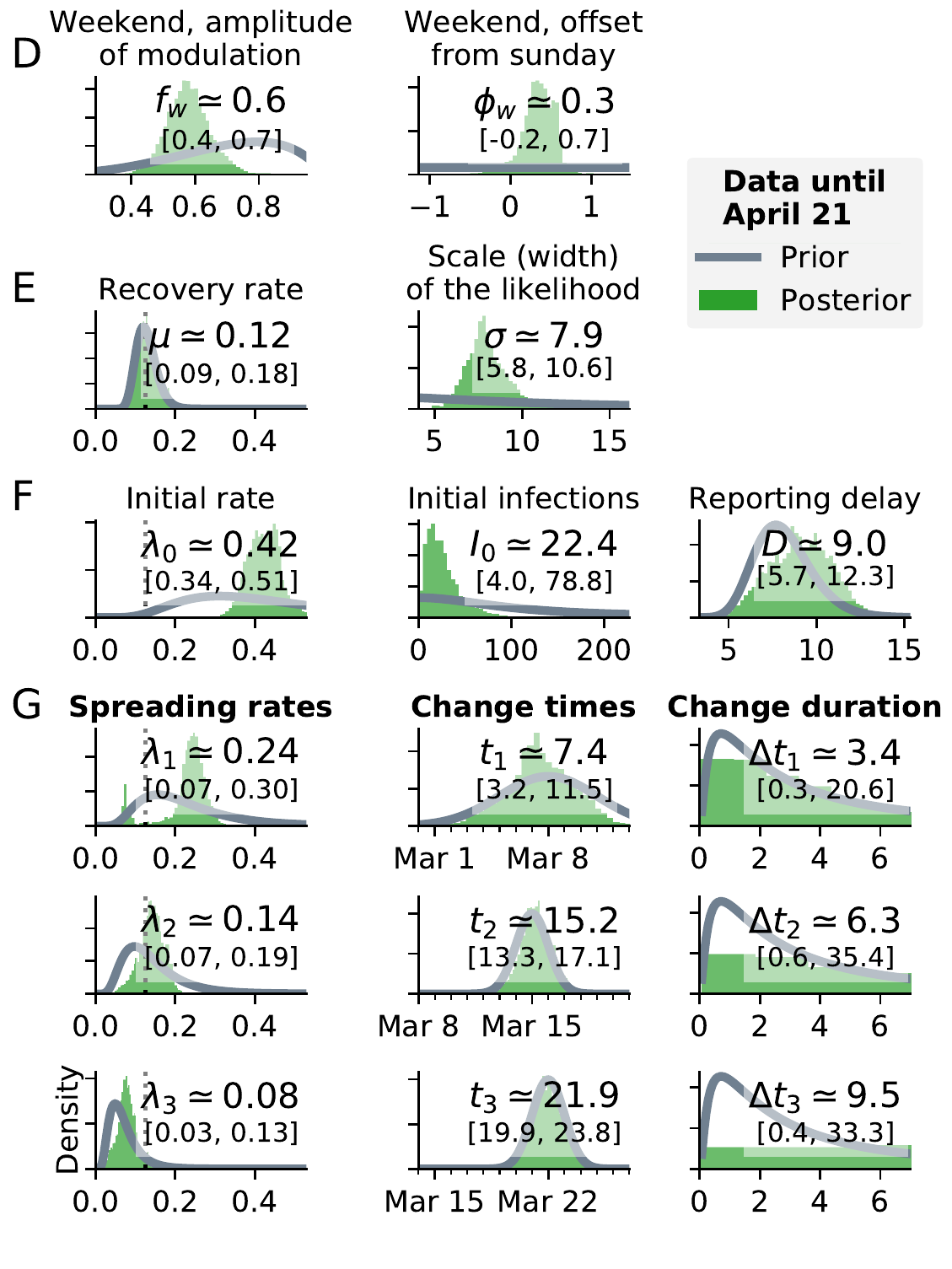}
    }
    \caption{
    \textbf{Sensitivity analysis:}
    Change-point detection as in Fig.~\ref{fig:inference_3_change_points} (main text, three change points, same model) but with \textbf{a prior for the change duration that is 4 times wider} (panel G, third column).
    The duration of the first change point $\Delta t_1$ is robust to the wider prior; the data constrains the posterior.
    The durations of the second and third change point, $\Delta t_2$ and $\Delta t_3$ are not constrained by the data but they depend on our chosen priors. This is also visible in the lack of plateaus in the effective growth, panel A.
    However, the inferred spreading rate $\lambda$ and the forecast are \textit{not sensitive} to the wider priors.
    \textbf{A:}~Time-dependent model estimate of the effective growth rate $\lambda^*(t)$. \textbf{B:}~Comparison of daily new reported cases and the model (inset: log-lin scale). \textbf{C:}~As B but for total (cumulative) reported cases. \textbf{D--G}:~Prior and posterior distributions of all free parameters.
    }
    \label{fig:SIR_wider_transient}
\end{figure}

\end{document}